\PassOptionsToPackage{dvipsnames}{xcolor}
\documentclass[11pt]{article}
\usepackage{jheppub}
\usepackage{bm,bbm}
\usepackage{booktabs,amsmath}
\usepackage{mathtools}
\usepackage{graphics}
\usepackage{braket}
\usepackage{tikz}
\usepackage{ascmac}
\usepackage{enumitem}
\usepackage{comment}
\usepackage{stackrel}
\usepackage{accents}
\usepackage{simpler-wick}
\usepackage{pb-diagram}

\everymath{\displaystyle}
\usepackage{cancel}

\usepackage{tabularx}
\renewcommand{\arraystretch}{1.3}
\usepackage{multirow}
\newcolumntype{M}[1]{>{\centering\arraybackslash}m{#1}}
\newcolumntype{N}{@{}m{0pt}@{}}
\usetikzlibrary{patterns}
\usetikzlibrary{decorations.markings}
\usetikzlibrary{decorations.pathmorphing}
\usetikzlibrary{shapes.geometric}
\usetikzlibrary{arrows}
\usetikzlibrary {arrows.meta}
\tikzset{snake it/.style={decorate, decoration=snake}}

\tikzset{test/.style n args={3}{
    postaction={
    decorate,
    decoration={
    markings,
    mark=between positions 0 and \pgfdecoratedpathlength step 0.5pt with {
    \pgfmathsetmacro\myval{multiply(
        divide(
        \pgfkeysvalueof{/pgf/decoration/mark info/distance from start}, \pgfdecoratedpathlength
        ),
        100
    )};
    \pgfsetfillcolor{#3!\myval!#2};
    \pgfpathcircle{\pgfpointorigin}{#1};
    \pgfusepath{fill};}
}}}}

\definecolor{c1}{rgb}{0.0,0.2,0.9}
\definecolor{c2}{rgb}{0.45,0.0,0.45}
\definecolor{c3}{rgb}{0.9,0.2,0.0}

\usetikzlibrary{arrows}

\tikzset{middlearrow/.style={
        decoration={markings,
            mark= at position 0.57 with {\arrow{#1}} ,
        },
        postaction={decorate}
    }
}

\usepackage{blkarray}

\def\ssymb#1{\mbox{\strut\rlap{\smash{\scriptsize$#1$}}\quad}}
\usepackage{nicematrix}
\NiceMatrixOptions%
{code-for-first-row = \scriptstyle,
code-for-first-col = \scriptstyle }

\usetikzlibrary{decorations.markings}
\def\MarkLt{4pt}
\def\MarkSep{2pt}

\tikzset{
  TwoMarks/.style={
    postaction={decorate,
      decoration={
        markings,
        mark=at position #1 with
          {
              \begin{scope}[xslant=0.2]
              \draw[line width=\MarkSep,white,-] (0pt,-\MarkLt) -- (0pt,\MarkLt) ;
              \draw[-] (-0.5*\MarkSep,-\MarkLt) -- (-0.5*\MarkSep,\MarkLt) ;
              \draw[-] (0.5*\MarkSep,-\MarkLt) -- (0.5*\MarkSep,\MarkLt) ;
              \end{scope}
          }
       }
    }
  },
  TwoMarks/.default={0.5},
  OneMark/.style={
    postaction={decorate,
      decoration={
        markings,
        mark=at position #1 with
          {
              \draw[-] (0,-\MarkLt) -- (0,\MarkLt) ;
          }
       }
    }
  },
  OneMark/.default={0.5}
}

\edef\restoreparindent{\parindent=\the\parindent\relax}
\usepackage{parskip}
\restoreparindent
\usepackage{ascmac}
\usepackage{mathrsfs}
\usepackage{enumitem}
\newlist{steps}{enumerate}{1}
\setlist[steps, 1]{label = Step \arabic*:}
\renewcommand*\arraystretch{1.2}

\usepackage{tikz}
\usetikzlibrary{intersections, calc,positioning,decorations.pathreplacing,decorations.pathmorphing}
\tikzset{>=latex}

\def\d{{\rm d}}

\def\i{{\rm i}}

\def\CB{{\cal B}}

\def\CD{{\cal D}}

\def\CL{{\cal L}}

\def\CN{{\cal N}}
\def\CO{{\cal O}}

\def\BH{\mathbb{H}}

\def\BR{\mathbb{R}}
\def\BS{\mathbb{S}}

\def\b0{\bm{0}_\perp}

\def\bP{\bold{P}}
\def\bM{\bold{M}}
\def\bD{\bold{D}}
\def\bK{\bold{K}}

\def\bJ{\bold{J}}

\def\SO{\mathrm{SO}}

\usepackage{centernot}
\usepackage{mathtools}
\usepackage{stmaryrd}

\makeatletter
\newcommand{\xMapsto}[2][]{\ext@arrow 0599{\Mapstofill@}{#1}{#2}}
\def\Mapstofill@{\arrowfill@{\Mapstochar\Relbar}\Relbar\Rightarrow}
\makeatother

\usepackage{tikz-3dplot}

\DeclareFontFamily{U}{mathx}{\hyphenchar\font45}
\DeclareFontShape{U}{mathx}{m}{n}{
      <5> <6> <7> <8> <9> <10>
      <10.95> <12> <14.4> <17.28> <20.74> <24.88>
      mathx10
      }{}
\DeclareSymbolFont{mathx}{U}{mathx}{m}{n}
\DeclareFontSubstitution{U}{mathx}{m}{n}
\DeclareMathAccent{\widecheck}{0}{mathx}{"71}
\title{Conformal field theory with composite defect}

\author[a]{Soichiro Shimamori}

\affiliation[a]{
Department of Physics, Osaka University,\\
Machikaneyama-Cho 1-1, Toyonaka 560-0043, Japan
}

\preprint{OU-HET-1231}

\abstract{We explore higher-dimensional conformal field theories (CFTs) in the presence of a conformal defect that itself hosts another sub-dimensional defect. We refer to this new kind of conformal defect as the \emph{composite defect}. We elaborate on the various conformal properties of the composite defect CFTs, including correlation functions, operator expansions, and conformal block expansions. As an example, we present a free O$(N)$ vector model in the presence of a composite defect. Assuming the averaged null energy condition (ANEC) does hold even for the defect systems, we conclude that some boundary conditions can be excluded. Our investigations shed light on the rich phenomenology arising from hierarchical defect structures, paving the way for a deeper understanding of critical phenomena in nature.
}

\begin{document}
\maketitle
\section{Introduction}
Conformal field theories (CFTs) have emerged as cornerstones in describing critical phenomena in our universe. They provide profound insights into the physical behaviors of systems at critical points, and play a crucial role in elucidating universal features that transcend microscopic details. However, our spacetime typically hosts defects whose shapes are distorted. Introducing these defects into spacetime leads to a loss of symmetries such as translation symmetries. Consequently, we cannot predict the dynamics of these defects in a controllable way. Nevertheless, non-trivial predictions can be made for such defect systems when spacetime possesses conformal symmetry and the shape of the defect is restricted to be planer or spherical. This kind of defect is referred to as the conformal defect, and the CFT in the presence of the conformal defect is to as defect CFT. Defect CFTs serve as indispensable tools for characterizing the dynamics of impurities, boundaries, or interfaces at critical points, ranging from condensed matter physics to high energy physics.
In recent years, studies on various conformal properties for a single conformal defect in $d$-dimensional CFT\footnote{Here, $d$ is greater than two.} have been deeply investigated in various directions e.g., kinematical structures \cite{Billo:2016cpy,Gadde:2016fbj,Lauria:2017wav, Lauria:2018klo,Kobayashi:2018okw,Guha:2018snh, Nishioka:2022ook}, dynamics of conformal defects \cite{McAvity:1995zd,Kapustin:2005py,Billo:2013jda, Gaiotto:2013nva, Yamaguchi:2016pbj, Herzog:2017xha, Herzog:2017kkj, Soderberg:2017oaa, Bissi:2018mcq, DiPietro:2019hqe, Herzog:2019bom, Prochazka:2019fah, Prochazka:2020vog, Lauria:2020emq, Herzog:2020lel, Herzog:2020bqw,Behan:2020nsf, DiPietro:2020fya, Cuomo:2021kfm, Cuomo:2021cnb, Bianchi:2021snj, Gimenez-Grau:2021wiv, Giombi:2021uae, Soderberg:2021kne, Behan:2021tcn, Collier:2021ngi, Ghosh:2021ruh, Bissi:2022bgu, Cuomo:2022xgw, Herzog:2022jqv, Herzog:2022jlx, Wang:2020xkc, Herzog:2023dop, Bartlett-Tisdall:2023ghh, SoderbergRousu:2023zyj, Giombi:2022vnz, Nishioka:2022odm, Nishioka:2022qmj, Gimenez-Grau:2022czc, Popov:2022nfq, Cuomo:2023qvp, Aharony:2023amq, DiPietro:2023gzi, Brax:2023goj, Dey:2024ilw}, renormalization group flows \cite{Jensen:2015swa, Casini:2018nym, Kobayashi:2018lil,Jensen:2018rxu, Giombi:2019enr, Cuomo:2021rkm, Wang:2021mdq, Nishioka:2021uef,Sato:2021eqo, Yuan:2022oeo,Shachar:2022fqk,Casini:2022bsu,Casini:2023kyj,Giombi:2023dqs, Raviv-Moshe:2023yvq, Trepanier:2023tvb, Zhou:2023fqu, Harper:2024aku}. 

\begin{figure}[t]
    \centering
       \begin{tikzpicture}
\draw[step=0.5cm,gray, very thick](-8,-13) grid (-4, -9);
\draw[orange!90, ultra thick] (-6, -13)--(-6, -9);
\foreach \x in {-8,-7.5,...,-6.5}
         {
             \foreach \y in {-13,-12.5,...,-9}
             {
                 \node[draw,circle,inner sep=1.5pt,fill] at (\x,\y) {};
             }
        }
 \foreach \x in {-5.5,-5,...,-4}
        {
            \foreach \y in {-13,-12.5,...,-9}
            {
                \node[draw,circle,inner sep=1.5pt,fill] at (\x,\y) {};
            }
       }
\foreach \y in {-13, -12.5, ..., -11.5}
{
    \node[draw,circle,inner sep=1.5pt,fill, orange!90] at (-6,\y) {};
}
\foreach \y in {-11, -10.5, ..., -9}
{
    \node[draw,circle,inner sep=1.5pt,fill, orange!90] at (-6,\y) {};
}
\node[draw,circle,inner sep=2pt,fill, teal] at (-6, -11) {};
\node at (-6, -14) {2D Quantum System};
\node at (3, -14) {(2+1)D Continuum System};
\fill[orange!50] (2, -13) --  (2, -9) -- (4, -8) -- (4, -12) --cycle;
\draw[ultra thick, teal] (3, -12.5)-- (3, -8.5);
\node[orange] at (1.5, -10) {$\widehat{\CD}^{(2)}$};
\node[teal] at (3.1, -13) {$\widecheck{\CD}^{(1)}$};
\node at (5.1, -7.7) {$\BR^{1,2}$};
\draw[thick] (4.7, -7.5)--(4.7, -8)--(5.2, -8);
\draw [arrows = {-Stealth[scale=1.5]}, blue!80, thick] (-3, -11)--(0, -11);
\node[blue!80, thick] at (-1.57, -10.5) {Continuum Limit};
\draw [arrows = {-Stealth[scale=1.1]}, thick] (5, -11)--(5, -9);
\node[thick] at (5, -8.7) {{\small time}};
\end{tikzpicture} 
    \caption{Our setup for composite defect CFTs. In the left panel, we illustrate a composite defect in a two-dimensional quantum system. In this system, a composite defect consists of an interface (orange line) that is contaminated by an impurity (teal point). In the right panel, we provide the schematic picture of the continuum theory which can be naively obtained by taking the limit of the left quantum system. In the continuum theory, an interface and an impurity become the worldsurface and worldline, respectively.}
    \label{fig: schematic picture of composite defect cft}
\end{figure}
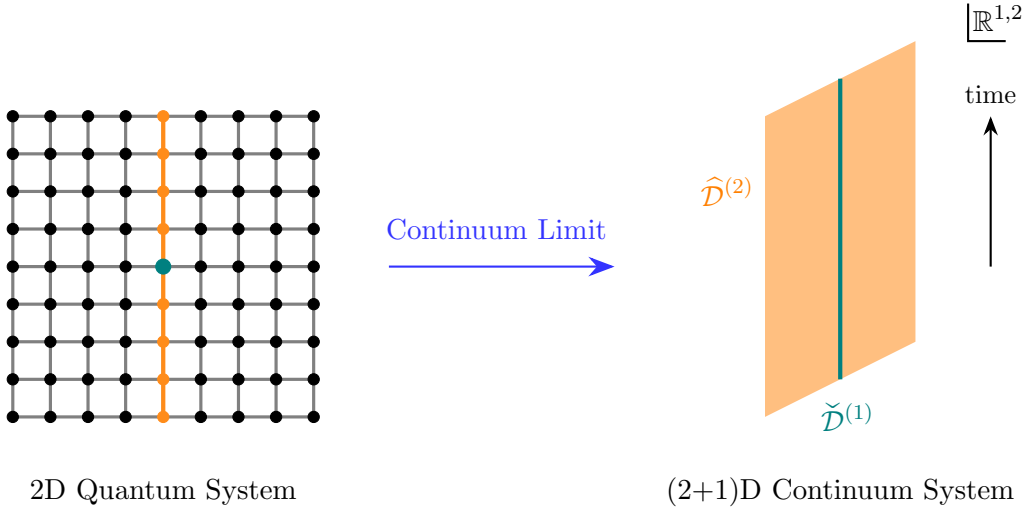
However, the real world often presents us with critical systems exhibiting hierarchical structures of defects, where a $p$-dimensional conformal defect $\widehat{\CD}^{(p)}$ itself hosts an additional $r$-dimensional one $\widecheck{\CD}^{(r)}$ with $r<p<d$. (In figure \ref{fig: schematic picture of composite defect cft}, we illustrate this setup both for two-dimensional quantum systems and three-dimensional continuum theory.) Throughout this paper, we refer to the former and latter defects as \emph{defect} and \emph{sub-defect}, respectively. Since this kind of defect can be conceptualized as a ``composite defect'' consisting of the defect $\widehat{\CD}^{(p)}$ and the sub-defect $\widecheck{\CD}^{(r)}$, we call the CFT in the presence of a composite defect as \emph{composite defect CFT}. We can naturally expect that the presence of a sub-defect gives rise to new universality classes that are unattainable within conventional defect CFTs and provides us with much richer phase structures of quantum field theories. 

In this paper, as a first step toward elucidating the conformal dynamics of composite defects, we develop the theoretical framework in defect CFTs to encompass such composite defect CFTs, particularly paying our attention to symmetry structures, conformal correlators, operator expansions, and conformal block expansions\footnote{A few remarks are in order for those who are familiar with wedge CFTs. If we restrict the defect dimensions such that $d=p+1=r+2$, the kinematical structures of the composite defect CFTs obtained in this paper is reduced to the ones of wedge CFTs which were developed in e.g., \cite{Cardy_1983, Antunes:2021qpy,Bissi:2022bgu}. We, however, stress that our underlying philosophy about the composite defect CFT is different from the wedge CFT. In wedge CFTs, we realize a co-dimension two defect (referred to as edge in the literature) by folding a boundary at some angle (see \cite[figure 1]{Antunes:2021qpy}). Therefore, the wedge CFTs become reduced to the ordinary boundary CFTs in the unfolding limit as described in \cite[section 2]{Antunes:2021qpy}. On the one hand, a composite defect is constructed by directly adding another dynamical sub-defect $\widecheck{\CD}^{(r)}$ to the defect $\widehat{\CD}^{(p)}$, hence we can realize non-trivial composite defect CFTs even in the case where the defect is planer.}. We also present the simplest example of the composite defect CFTs: free O$(N)$ vector model. For this model, we can assign Neumann or Dirichlet boundary conditions of a bulk fundamental field $\phi^{\, I} (I=1, 2, \cdots, N)$ for two kinds of defects $\widehat{\CD}^{(p)}$ and $\widecheck{\CD}^{(r)}$. We explore various conformal properties of this concrete model including the operator spectrums of a composite defect, operator expansions, etc. We also show that some boundary conditions can be excluded by using the constraints from the averaged null energy condition (ANEC).

This paper is structured as follows. In section \ref{sec: kinematical constraint}, we elaborate on the various model-independent properties of composite defect CFTs. 
In section \ref{subsec: residual symmetry}, we discuss the residual conformal symmetry in the presence of a composite defect. In section \ref{subsec: correlator}, we explain how we can determine the form of scalar correlation functions in composite defect CFTs only from the residual symmetry. In section \ref{subsec: OE}, we discuss the operator expansions appearing in composite defect CFTs. We particularly show that a bulk local field can be expanded in terms of sub-defect local ones. In section \ref{subsec: conformal blocks}, we derive the conformal block expansions by focusing on a bulk one-point function and a bulk--sub-defect two-point one. In section \ref{subsec: O(N) free scalar model}, we introduce the O($N$) vector model which is the simplest model of the composite defect CFTs. In particular, we characterize a composite defect by the boundary conditions of a bulk field for a defect $\widehat{\CD}^{(p)}$ and $\widecheck{\CD}^{(r)}$. In section \ref{subsec: defect CFT data} and \ref{subsec: subdefect data}, we describe the defect and sub-defect operator spectrums, respectively. We also derive the exact formulas of sub-defect operator expansions for a bulk field depending on its boundary conditions. In section \ref{subsec: em tensor}, we derive the energy-momentum tensor in the O$(N)$ vector model, from which in section \ref{subsubsec: ANEC constraint}, we argue that some boundary conditions can be excluded by assuming that the ANEC does hold even in the defect systems. In section \ref{sec: conclusion and future}, we summarize this paper and describe some future directions. In appendix \ref{sec: method of images scalar}, we discuss the method of images in composite defect CFTs. In appendix \ref{appendix: proof of OEs}, we provide the proof for the selected operator expansion which is omitted in the main text.

\section{Composite defect CFT}\label{sec: kinematical constraint}
In this section, we discuss some conformal aspects of composite defect CFTs. We emphasize that the contents of this section can be applied to arbitrary composite defect CFTs. In section \ref{subsec: residual symmetry}, we explain how the full conformal symmetry in $d$-dimensional spacetime is broken into its subgroup by employing conformal maps. In section \ref{subsec: correlator}, we determine the form of correlation functions by making use of the residual symmetry discussed in section \ref{subsec: residual symmetry}. In section \ref{subsec: OE}, we speculate on the operator expansions in composite defect CFTs. Finally, in section \ref{subsec: conformal blocks}, we derive differential equations that conformal blocks must satisfy, and give some analytical expressions of conformal blocks. 
\subsection{Symmetry in composite defect CFT}\label{subsec: residual symmetry}
\begin{figure}[t]
    \centering
       \begin{tikzpicture}
        \coordinate (A) at (0 , 0) {};
        \coordinate (B) at (5 , 0) {};
        \coordinate (C) at (6 , 2) {};
        \coordinate (D) at (1 , 2) {};
        \draw[thick, black!100,->,>=stealth] (7, 2.5)  -- (7.3, 3.1)  node[right, black] {$\hat{x}^2$};
        \draw[thick, black!100,->,>=stealth] (7, 2.5)  -- (7, 3.5) node[above, black] {$x_{\perp}^3$};
        \draw[thick, black!100,->,>=stealth] (7, 2.5)  -- (8, 2.5) node[right, black] {$\check{x}^1$};
        \fill[orange!50] (A)--(B)--(C)--(D)--cycle;
        \draw[ultra thick, teal] ($(A)!0.5!(D)$) -- ($(B)!0.5!(C)$);
        \node[left=0.2cm, teal] at ($(A)!0.5!(D)$) {$\widecheck{\CD}^{(1)}$};
        \node[below=0.2cm, orange] at ($(A)!0.5!(B)$) {$\widehat{\CD}^{(2)}$};
     \end{tikzpicture} 
    \caption{Illustrative example for a composite defect system. In this figure, we consider $d=3$ bulk spacetime where a composite defect $\CD^{(2, 1)}\equiv \widehat{\CD}^{(2)}\,\cup \,\widecheck{\CD}^{(1)}$ is inserted. Now, a three dimensional bulk coordinate is given by $x^{\mu}=(\check{x}^{1}, \hat{x}^{2}, x^{3}_{\perp})$, where a two-dimensional defect $\widehat{\CD}^{(2)}$ sits at $x^{3}_{\perp}=0$, and a one-dimensional defect $\widecheck{\CD}^{(1)}$ is stretched in the direction of the $\check{x}^{1}$ axis lying on $\widehat{\CD}^{(2)}$.}
    \label{fig: composite defect system}
\end{figure}
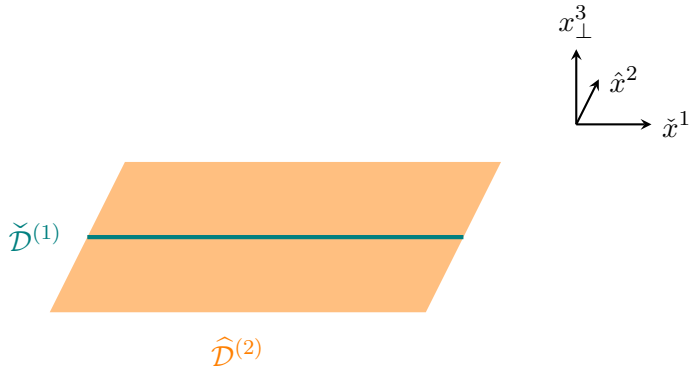
In this subsection, we investigate the residual conformal symmetry in composite defect CFTs. We consider a CFT in $d$-dimensional spacetime $X^{d}$ whose coordinate is described by $x^{\mu}$ ($\mu=1, 2, \cdots , d$), and assume the existence of a composite defect $\CD^{(p, r)}$:
\begin{align}
    \CD^{(p, r)}\equiv \widehat{\CD}^{(p)}\,\cup \,\widecheck{\CD}^{(r)}\qquad , \qquad r<p<d\quad , 
\end{align}
in a bulk spacetime. Here, $\widehat{\CD}^{(p)}$ is a $p$-dimensional conformal (i.e., planer or spherical) defect, and $\widecheck{\CD}^{(r)}$ is a $r$-dimensional conformal sub-defect which is embedded into $\widehat{\CD}^{(p)}$. For simplicity, we confine ourselves to the planar composite defect, and set its location as follows:
\begin{align}\label{eq: location of composite defect}
    \begin{aligned}
        \widehat{\CD}^{(p)}\ &: \ \ \ \check{x}^{a}\not=0\quad , \quad \hat{x}^{\alpha}\not=0\quad , \quad x^{i}_{\perp}=0\ ,  \\
        \widecheck{\CD}^{(r)}\ &: \ \ \ \check{x}^{a}\not=0\quad , \quad \hat{x}^{\alpha}=0\quad , \quad x^{i}_{\perp}=0\ , 
    \end{aligned}
\end{align}
where the coordinate system is 
\begin{align}
    x^{\mu}=(\check{x}^{a}, \hat{x}^{\alpha}, x^{i}_{\perp})\quad , \quad a=1\sim r\quad , \quad \alpha=r+1\sim p \quad , \quad i=p+1\sim d \ . 
\end{align}
(In figure \ref{fig: composite defect system}, we provide an illustrative example that helps the readers with imagining our setup.)

A $d$-dimensional (Euclidean) CFT possesses the conformal symmetry $\text{SO}(1, d+1)$. Introducing some conformal defect typically breaks $\text{SO}(1, d+1)$ symmetry to its subgroup. If we introduce a composite defect $\CD^{(p, r)}$ into a spacetime, the full conformal symmetry is broken as follows:
 \begin{align}\label{eq: residual symmetry}
     \text{insertion of }\CD^{(p, r)}\ : \ \text{SO}(1, d+1)\longrightarrow \text{SO}(1, r+1)\times \text{SO}(p-r)\times \text{SO}(d-p)\ , 
 \end{align}
 where $\SO(1, r+1)$ is the conformal symmetry along with the sub-defect $\widecheck{\CD}^{(r)}$, and $\SO(d-p)$ and $\SO(p-r)$ are rotational symmetries.
\begin{figure}[t]
    \centering
       \begin{tikzpicture}
\fill[orange!50] (-10.5, -3) --  (-10.5, 1) -- (-8.5, 2) -- (-8.5, -2) --cycle;
\draw[ultra thick, teal] (-9.5, -2.5)-- (-9.5, 1.5);
\node[orange] at (-11, 0) {$\widehat{\CD}^{(p)}$};
\node[teal] at (-9.4, -3) {$\widecheck{\CD}^{(r)}$};
\node at (-7.45, 2.3) {$\BR^d$};
\draw[thick] (-7.8, 2.5)--(-7.8, 2)--(-7.3, 2);

\node at (-7, -0.5) {{\huge $\cong$}};

\node[ellipse, minimum width=1cm, minimum height=3cm, outer sep=0, label=180:] (ell1) at (-4.5,-0.5) {};

\fill[orange!50] (ell1.85) to[out=-45,in=175] (-2,-0.3)
to[out=-45, in=45] (-2,-0.5)
to[out=-175, in=45] (ell1.-85);
\node at (-3.5,-2.7) {$\BH^{r+1}$};

\node[ellipse, fill=teal!60, minimum width=1cm, minimum height=3cm, outer sep=0, label=180:\textcolor{teal}{$\widecheck{\CD}^{(r)}$}] (ell) at (-4.5,-0.5) {};

\node at (-1.5, -0.5) {\Large $\times$};

\draw[thick] (-0.5, -2.5) -- (3.5, -2.5);
\fill[black] (-0.5, -2.5) circle [radius=0.1];
\node[below=0.2cm] at (-0.5, -2.5) {$\theta=0$};
\fill[black] (3.5, -2.5) circle [radius=0.1];
\node[below=0.2cm] at (3.5, -2.5) {$\theta=\pi/2$};

 \draw[thick] (-0.5, 2)--( 3.5, 1)--(-0.5, 0);
 \fill[orange!50] (-0.5 , 1) ellipse (0.4cm and 1cm);

 \draw[thick] (3.5, 0)--( -0.5, -1)--(3.5, -2);
 \draw[thick] (3.5 , -1) ellipse (0.4cm and 1cm);
 \fill[orange!50] (-0.5, -1) circle[radius=0.1];

\node at (-1.7,1.3) {$\BS^{p-r-1}$};

\node at (4.8,-1) {$\BS^{d-p-1}$};

\end{tikzpicture} 
    \caption{Flat space $\BR^{d}$ with a composite defect $\widehat{\CD}^{(p, r)}$ is conformally equivalent ($\,\cong\,$) to the product of the $(r+1)$-dimensional Euclidean AdS spacetime $\BH^{r+1}$, and the spherical geometry which is characterized by \eqref{eq: spherical geometry}. Particularly, the sub-defect $\widecheck{\CD}^{(r)}$ is mapped to the AdS boundary while the defect $\widehat{\CD}^{(p)}$ is to some $(r+1)$-dimensional sub-manifold of $\BH^{r+1}$ and $(p-r-1)$-dimensional unit sphere $\BS^{p-r-1}$ at $\theta=0$.}
    \label{fig: conformal map}
\end{figure}
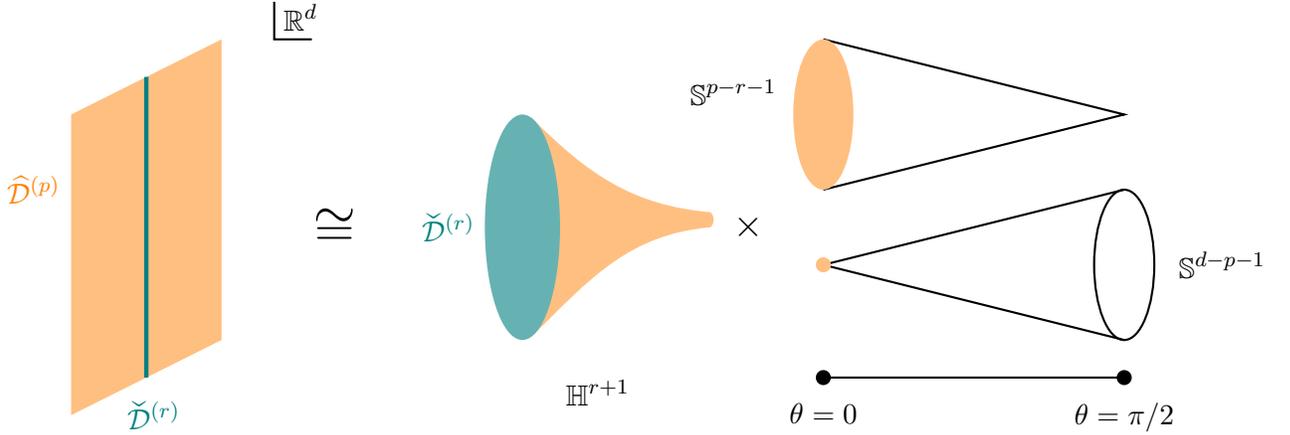

 We here give an interpretation of this residual symmetry \eqref{eq: residual symmetry} from the viewpoint of the conformal map along with \cite{Jensen:2013lxa} (see also e.g., \cite{Kapustin:2005py, Lewkowycz:2013laa, Giombi:2021uae}). In our composite defect systems, we have the following flat metric:
\begin{align}\label{eq: flat metric}
    \d s^{2}=\sum_{a=1}^{r}\d \check{x}_{a}^{2}+\sum_{\alpha=r+1}^{p}\d \hat{x}_{\alpha}^{2}+\sum_{i=p+1}^{d}\d x_{\perp, i}^{2}\ , 
\end{align} 
where a composite defect $\CD^{(p, r)}$ is located as \eqref{eq: location of composite defect}. We first move on to the cylinder coordinates:
\begin{align}
    \sum_{\alpha=r+1}^{p}\d \hat{x}_{\alpha}^{2}=\d \hat{z}^{2}+\hat{z}^{2}\, \d \Omega^{2}_{p-r-1}\quad , \quad \sum_{i=p+1}^{d}\d x_{\perp, i}^{2}=\d z_{\perp}^{2}+z_{\perp}^{2}\, \d \Omega^{2}_{d-p-1}\ , 
\end{align}
where $\hat{z}, z_{\perp}\in[0, \infty)$ denote the radial coordinates, and $\d\Omega_{q}^{2}$ is the metric for the $q$-dimensional unit sphere $\BS^{q}$. After performing this coordinate transformation, the flat metric \eqref{eq: flat metric} becomes
\begin{align}\label{eq: intermediate metric}
    \d s^{2}=\sum_{a=1}^{r}\d \check{x}_{a}^{2}+\d \hat{z}^{2}+\d z_{\perp}^{2}+\hat{z}^{2}\, \d \Omega^{2}_{p-r-1}+z_{\perp}^{2}\, \d \Omega^{2}_{d-p-1}\ . 
\end{align}
We moreover transfer from the $(\hat{z}, z_{\perp})$ coordinates to the polar ones $(z, \theta)$: 
\begin{align}
    \hat{z}=z\, \cos\theta\quad , \quad z_{\perp}=z\, \sin\theta\ , 
\end{align}
where $y\geq 0$ and $0\leq\theta\leq\pi/2$, and then the metric \eqref{eq: intermediate metric} becomes transformed into
\begin{align}\label{eq: conf equiv metric}
    \d s^{2}=z^{2}\left(\d s^{2}_{\BH^{r+1}}+\d s^{2}_{\text{sph}}\right)\ .
\end{align}
Here, $\d s^{2}_{\BH^{r+1}}$ is the metric for the $(r+1)$-dimensional Euclidean AdS spacetime $\BH^{r+1}$:
\begin{align}\label{eq: ads metric}
    \d s^{2}_{\BH^{r+1}}=\frac{\sum_{a=1}^{r}\d \check{x}_{a}^{2}+\d z^{2}}{z^{2}}\ , 
\end{align}
and the sub-defect $\widecheck{\CD}^{(r)}$ is located at the AdS boundary (namely, $z=0$). Also, $\d s^{2}_{\text{sph}}$ is the spherical metric which is defined by:
\begin{align}\label{eq: spherical geometry}
    \d s^{2}_{\text{sph}}=\d \theta^{2}+\cos^{2}\theta\, \d \Omega^{2}_{p-r-1}+\sin^{2}\theta\, \d \Omega^{2}_{d-p-1}\ .
\end{align}
For a fixed value of $\theta$, the above spherical geometry consists of the $(p-r-1)$-dimensional sphere whose radius is $\cos\theta$ and the $(d-p-1)$-dimensional one whose radius is $\sin\theta$. Therefore, by applying the Weyl rescaling to \eqref{eq: conf equiv metric}, we can convince that the flat spacetime in the presence of a composite defect $\CD^{(p, r)}$ is conformally equivalent to the product of the $(r+1)$-dimensional Euclidean AdS spacetime \eqref{eq: ads metric} and the spherical geometry \eqref{eq: spherical geometry}. We illustrate this picture in figure \ref{fig: conformal map}. We should notice that the metric \eqref{eq: conf equiv metric} has the isometry $\text{SO}(1, r+1)\times \text{SO}(p-r)\times \text{SO}(d-p)$ which is exactly coincident with the residual conformal symmetry \eqref{eq: residual symmetry}. This provides one way to understand the residual conformal symmetry for composite defect CFTs.

Finally, it is instructive to discuss which symmetry generators of $\text{SO}(1, d+1)$ do survive in the presence of a composite defect $\CD^{(p, r)}$. The symmetry generators of the full conformal symmetry $\text{SO}(1, d+1)$ are described by $\mathbf{J}_{MN}$ $(M, N=-1, 0, 1, \cdots, d)$ satisfying the following commutation relations:
\begin{align}\label{eq: JMN Euclid}
    \begin{aligned}
        &[\bJ_{MN}, \bJ_{RS}]=-\eta_{MR}\, \bJ_{NS}+\eta_{NR}\, \bJ_{MS}+\eta_{MS}\, \bJ_{NR}-\eta_{NS}\, \bJ_{MR} \ , \\
        &\eta_{MN}\equiv\text{diag}(-1, 1,1, \cdots, 1)\ . 
    \end{aligned}
\end{align}
The relations between the generators $\bJ_{MN}$ and the dilatation, translation, rotation, and special conformal transformation generators $\{\bD, \bP_{\mu}, \bM_{\mu\nu}, \bK_{\mu}\}$ are given by
\begin{align}\label{eq:conformal group embedding Euclid}
\bJ_{MN}
  = 
   \begin{pNiceMatrix}[first-row,first-col]
        \ssymb{\substack{\ \\ \ \\ M\backslash\, \\\downarrow}} \ssymb{\substack{ N\to\\ \, }} & -1 & 0 & \nu \\
        -1 & 0 & \bD & \frac{1}{2}(\bP_\nu-\bK_\nu) \\
        0 & -\bD& 0 & \frac{1}{2}(\bP_\nu+\bK_\nu) \\
        \mu &-\frac{1}{2}(\bP_\mu-\bK_\mu)& - \frac{1}{2}(\bP_\mu+\bK_\mu) & \bM_{\mu\nu}
    \end{pNiceMatrix}\ .
\end{align}
Clearly from the symmetry breaking pattern \eqref{eq: residual symmetry}, the symmetry generators for the residual symmetry $\text{SO}(1, r+1)\times\text{SO}(p-r)\times \text{SO}(d-p)$ are block-diagonal parts of $\bJ_{MN}$ as follows:
\begin{align}\label{eq:defect conformal group embed}
  \mathbf{J}_{MN}= \begin{pNiceMatrix}[first-row,first-col]
\ssymb{\substack{\ \\ \ \\ M\backslash\, \\\downarrow}} \ssymb{\substack{ N\to\\ \, }} & -1 & 0 & b & \beta & j \\
-1 & 0 & \mathbf{D} & \frac{1}{2}(\mathbf{P}_b-\mathbf{K}_b)& \cancel{\frac{1}{2}(\mathbf{P}_\beta-\mathbf{K}_\beta)} &\cancel{\frac{1}{2}(\mathbf{P}_j-\mathbf{K}_j)} \\
0 & -\mathbf{D}& 0 & \frac{1}{2}(\mathbf{P}_b+\mathbf{K}_b) & \cancel{\frac{1}{2}(\mathbf{P}_\beta+\mathbf{K}_\beta)} & \cancel{\frac{1}{2}(\mathbf{P}_j+\mathbf{K}_j)}\\
a &-\frac{1}{2}(\mathbf{P}_a-\mathbf{K}_a)& - \frac{1}{2}(\mathbf{P}_a+\mathbf{K}_a) & \mathbf{M}_{ab} &\cancel{\mathbf{M}_{a\beta}} &\cancel{\mathbf{M}_{aj}} \\
\alpha &  \cancel{-\frac{1}{2}(\mathbf{P}_\alpha-\mathbf{K}_\alpha)}& \cancel{- \frac{1}{2}(\mathbf{P}_\alpha+\mathbf{K}_\alpha)} & \cancel{\mathbf{M}_{\alpha b}} & \mathbf{M}_{\alpha\beta} & \cancel{\mathbf{M}_{\alpha j}} \\
i &  \cancel{-\frac{1}{2}(\mathbf{P}_i-\mathbf{K}_i)}& \cancel{- \frac{1}{2}(\mathbf{P}_i+\mathbf{K}_i)} & \cancel{\mathbf{M}_{ib}} & \cancel{\mathbf{M}_{i\beta}} & \mathbf{M}_{ij}
\end{pNiceMatrix}\ , 
\end{align}
where the slashed generators correspond to the broken ones by the insertion of $\CD^{(p, r)}$.

In composite defect CFTs, the number of symmetries is of course fewer compared to homogeneous CFTs or conventional defect CFTs. Therefore, it is non-trivial to determine how much restrictions can be imposed on composite defect CFTs only from the residual symmetry. In the rest of section \ref{sec: kinematical constraint}, we address this issue by focusing on conformal correlators (section \ref{subsec: correlator}), operator expansions (section \ref{subsec: OE}), and conformal blocks (section \ref{subsec: conformal blocks}).

\subsection{Scalar correlators in composite defect CFTs}\label{subsec: correlator}
In this section, we elaborate on how the correlation functions can be fixed only from the residual conformal symmetry $\text{SO}(1, r+1)\times \text{SO}(p-r)\times \text{SO}(d-p)$. In composite defect CFTs, local operators are classified into the following three classes:
\begin{itemize}
    \item Bulk local operator at $x^{\mu}\equiv(\check{x}^{a}, \hat{x}^{\alpha}, x_{\perp}^{i})$: 
    \begin{align}\label{eq: bulk local op}
        \CO_{\Delta, J}(x) \ , 
    \end{align}
    where $\Delta$ is the conformal dimension and $J$ is $\text{SO}(d)$ spin number. (In this expression, we omit tensorial indices to avoid some complicated notations. This remark is also applied to \eqref{eq: defect local op} and \eqref{eq: sub-defect local op}.)
    \item Defect local operator at $\hat{\bm{x}}^{\mu}\equiv (\check{x}^{a}, \hat{x}^{\alpha}, x_{\perp}^{i}=0)$: 
    \begin{align}\label{eq: defect local op}
        \widehat{\CO}_{\widehat{\Delta}, (\ell, s)}\,(\hat{\bm{x}})\ ,
    \end{align} 
    where $\widehat{\Delta}$ is the conformal dimension, and $\ell$ and $s$ are the $\text{SO}(p)$ and $\text{SO}(d-p)$ spin numbers, respectively.
    \item Sub-defect local operator at $\check{\bm{x}}^{\mu}\equiv(\check{x}^{a}, \hat{x}^{\alpha}=0, x_{\perp}^{i}=0)$: 
    \begin{align}\label{eq: sub-defect local op}
        \widecheck{\CO}_{\widecheck{\Delta}, (\ell, s_{1}, s_{2})}(\check{\bm{x}})\ , 
    \end{align}
    where $\widecheck{\Delta}$ is the conformal dimension, and $\ell$, $s_{1}$ and $s_{2}$ are the $\text{SO}(r)$, $\text{SO}(p-r)$ and $\text{SO}(d-p)$ spin numbers, respectively.
\end{itemize}
In this subsection, we argue how the residual symmetry \eqref{eq: residual symmetry} puts restrictions on the conformal correlators composing scalar primaries.
\subsubsection{Preliminaries: embedding space formalism}\label{subsubsec: review on embedding space formalism}
 Our methodology to fix the conformal correlators relies on the \emph{embedding space formalism}, which has been developed in \cite{Dirac:1936fq, Weinberg:2010fx, Costa:2011mg}. In this subsection, we present a brief review of the embedding space formalism in the absence of any defects. (We recommend e.g. \cite{Costa:2011mg, Rychkov:2016iqz, Penedones:2016voo} for the readers who want to know more details.) If readers are familiar with the embedding space formalism in $d$-dimensional CFT, you may skip this subsection and move on to the next one.

In the embedding space formalism, we lift up the $d$-dimensional Euclidean spacetime $\BR^{d}$ to the projective null cone $\mathbb{PNC}^{1, d+1}$ which is defined by
\begin{align}
    \mathbb{PNC}^{1, d+1}\equiv\frac{\{X^{M}\in \BR^{1, d+1}; X\cdot X\equiv \eta_{MN}X^{M}X^{N}=0\}}{\{X^{M}\sim \lambda\, X^{M}; \forall\lambda>0\}}\ , 
\end{align}
where $\eta_{MN}$ is the Lorentzian metric given in \eqref{eq: JMN Euclid}, and $X^{M}\sim\lambda X^{M}$ means that we identify $X^{M}$ up to positive scalar multiplication. On the projective null cone $\mathbb{PNC}^{1, d+1}$, the conformal transformation acts linearly on the embedding space coordinate $X^{M}$:
\begin{align}
    X^{M}\mapsto \Lambda^{M}_{N}\, X^{N}\qquad , \qquad \Lambda^{M}_{N}\in \text{SO}(1, d+1)\ , 
\end{align}
and the scalar primary operator $\CO_{\Delta}(X)$ satisfies
\begin{align}\label{eq: homogeneity condition}
    \CO_{\Delta}(\lambda X)=\lambda^{-\Delta}\CO_{\Delta}(X)\ . 
\end{align}
While the conformal transformations in the physical space non-linearly act on fields, those in the embedding space formalism does linearly. This fact makes it much easier to construct conformal correlators.   
One way to get back to the physical space $\BR^{d}$ is to restrict the embedding space coordinate $X^{M}$ as follows:
\begin{align}\label{eq: Poincare section}
    \left.X^{M}\right|_{\text{phys}}=\left(\frac{x^{2}+1}{2}, \frac{x^{2}-1}{2}, x^{\mu}\right)\ .
\end{align}
Under this restriction, a scalar primary operator $\CO_{\Delta}(X)$ on the projective null cone $\mathbb{PNC}^{1, d+1}$ is reduced to the one $\CO_{\Delta}(x)$ on the physical space:
\begin{align}\label{eq: field on phys space}
    \CO_{\Delta}(x)=\CO_{\Delta}\left(X|_{\text{phys}}\right)\ . 
\end{align}

Here, we give an example: the scalar two-point function $\left\langle\,\CO_{\Delta_{1}}(x_{1})\CO_{\Delta_{2}}(x_{2})\right\rangle_{\BR^{d}}$. We first need to lift its correlator on the physical space $\BR^{d}$ to the one on the projective null cone $\mathbb{PNC}^{1, d+1}$:
\begin{align}
    \left\langle\CO_{\Delta_{1}}(x_{1})\CO_{\Delta_{2}}(x_{2})\right\rangle_{\BR^{d}}\longrightarrow \left\langle\CO_{\Delta_{1}}(X_{1})\CO_{\Delta_{2}}(X_{2})\right\rangle_{\mathbb{PNC}^{1, d+1}}\ . 
\end{align}
Now that we have the full conformal symmetry $\text{SO}(1, d+1)$, the conformal correlator should consist of the SO($1, d+1$) invariant quantities. In this case, $X_{1}\cdot X_{2}$ is the clearly unique invariant up to scalar multiplications due to null properties: $X^{2}_{1}=X^{2}_{2}=0$. Therefore, the two-point function should be some function of $X_{1}\cdot X_{2}$:
\begin{align}
    \left\langle\CO_{\Delta_{1}}(X_{1})\CO_{\Delta_{2}}(X_{2})\right\rangle_{\mathbb{PNC}^{1, d+1}}=f(X_{1}\cdot X_{2})\ . 
\end{align}
From the homogeneity condition \eqref{eq: homogeneity condition}, the function $f(X_{1}\cdot X_{2})$ should satisfy
\begin{align}
    f(\lambda X_{1}\cdot X_{2})=\lambda^{-\Delta_{1}}f(X_{1}\cdot X_{2})=\lambda^{-\Delta_{2}}f(X_{1}\cdot X_{2})\ , 
\end{align}
from which we can fix the form of the two-point function as follows:
\begin{align}
    \left\langle\CO_{\Delta_{1}}(X_{1})\CO_{\Delta_{2}}(X_{2})\right\rangle_{\mathbb{PNC}^{1, d+1}}=\frac{c\, \delta_{\Delta_{1}, \Delta_{2}}}{(-2X_{1}\cdot X_{2})^{\Delta_{1}}}\ , 
\end{align}
where $c$ is some constant, and $-2$ appearing in front of $X_{1}\cdot X_{2}$ is just for later convenience. Finally, in order to come back to the physical space, we need to restrict the embedding space coordinate onto \eqref{eq: Poincare section}. By using \eqref{eq: field on phys space} and the following relation:
\begin{align}
    \left.-2X_{1}\cdot X_{2}\right|_{\text{phys}}=|x_{12}|^{2}\qquad , \qquad x_{12}\equiv x_{1}-x_{2}\ , 
\end{align}
we arrive at 
\begin{align}
    \left\langle\CO_{\Delta_{1}}(x_{1})\CO_{\Delta_{2}}(x_{2})\right\rangle_{\BR^{d}}=\frac{c\, \delta_{\Delta_{1}, \Delta_{2}}}{|x_{12}|^{2\Delta_{1}}}\ , 
\end{align}
which is a well-known result in $d$-dimensional CFT. 
\subsubsection{Scalar correlators in composite defect CFT}\label{subsubsec: Scalar correlators in composite defect CFT}
In the previous subsection, we briefly reviewed the embedding space formalism. In this subsection, we apply the embedding space formalism to composite defect CFTs, and explore conformal structures of correlators consisting of only scalar primaries.

As discussed in subsection \ref{subsec: residual symmetry}, we have the residual symmetry $\text{SO}(1, r+1)\times \text{SO}(p-r)\times \text{SO}(d-p)$. It is therefore convenient to introduce three kinds of inner products $\bullet$, $\diamond$ and $\circ$ which are defined as follows:
\begin{align}
    X\bullet Y&\equiv\eta_{AB}X^{A}Y^{B}\quad , \quad A, B=-1, 0, 1, \cdots , r\ , \\
     X\diamond Y&\equiv\delta_{\alpha\beta}X^{\alpha}Y^{\beta}\quad , \quad \alpha, \beta=r+1, r+2, \cdots , p\ , \\
     X\circ Y&\equiv\delta_{ij}X^{i}Y^{j}\quad , \quad i, j=p+1, r+2, \cdots , d\ .
\end{align}
These three inner products are invariant under the residual symmetry transformations and play a crucial role in the construction of the conformal correlators in composite defect CFTs.
\paragraph{Bulk one-point function.} We first discuss the bulk one-point function $\langle\, \CO_{\Delta}(x)\, \rangle$. In the same way as the subsection \ref{subsubsec: review on embedding space formalism}, we promote it to the correlator on the projective null cone $\mathbb{PNC}^{1, d+1}$, and we consider $\langle\, \CO_{\Delta}(X)\, \rangle$\footnote{For simplicity, in the rest of this paper, we leave the symbols $\BR^{d}$ and $\mathbb{PNC}^{1, d+1}$ out of correlators.}. By noting the null condition:
\begin{align}
    X\cdot X=X\bullet X+X\diamond X+X\circ X=0\ ,
\end{align}
we can construct two independent invariants:
\begin{align}
    X\diamond X\qquad , \qquad X\circ X\ , 
\end{align}
from which we can also consider the cross-ratio $\xi$ defined by
\begin{align}\label{eq: cross-ratio 1}
    \xi\equiv \frac{X\circ X}{X\diamond X} \ . 
\end{align}
Thanks to the homogeneity condition \eqref{eq: homogeneity condition}, the bulk one-point function in the embedding space can be fixed as
\begin{align}\label{eq: bulk one pt}
    \langle\, \CO_{\Delta}(X)\, \rangle=\frac{f(\xi)}{(X\circ X)^{\Delta/2}}\ , 
\end{align}
where $f(\xi)$ is some function depending only on the cross-ratio $\xi$. In the physical space, the cross-ratio $\xi$ and the inner product $X\circ X$ become
\begin{align}
    \xi\longrightarrow \frac{|x_{\perp}|^2}{|\hat{x}|^{2}}\qquad , \qquad X\circ X\longrightarrow |x_{\perp}|^{2}\ , 
\end{align}
therefore the bulk one-point function in the physical space is given by\footnote{In the following, we employ the same symbols for both cross-ratios defined in physical and embedding space. We believe that readers will not be confused about this notation.}
\begin{align}
    \langle\, \CO_{\Delta}(x)\, \rangle=\frac{f(\xi)}{|x_{\perp}|^{\Delta}}\qquad , \qquad \xi=\frac{|x_{\perp}|^2}{|\hat{x}|^{2}}\ .  
\end{align}
\paragraph{Defect one-point function.} We next consider the defect one-point function $\langle\, \widehat{\CO}_{\widehat{\Delta}}(\hat{\bm{x}})\,\rangle$ where $\hat{\bm{x}}^{\mu}=(\check{x}^{a}, \hat{x}^{\alpha}, x_{\perp}^{i}=0)$. The corresponding correlator in the embedding space is given by
\begin{align}
    \langle\, \widehat{\CO}_{\widehat{\Delta}}(\widehat{X}\,)\,\rangle \quad , \quad \widehat{X}\cdot\widehat{X}=\widehat{X}\bullet\widehat{X}+\widehat{X}\diamond\widehat{X}=0\ . 
\end{align}
In this case, there is only one invariant $\widehat{X}\diamond \widehat{X}$, and the homogeneity condition fixes the form of the defect one-point function:
\begin{align}
     \langle\, \widehat{\CO}_{\widehat{\Delta}}(\widehat{X}\,)\,\rangle =\frac{c}{(\widehat{X}\diamond \widehat{X})^{\widehat{\Delta}/2}}\ , 
\end{align}
where $c$ is some constant. By the reduction to the physical space, we have
\begin{align}
    \langle\, \widehat{\CO}_{\widehat{\Delta}}(\hat{\bm{x}})\,\rangle=\frac{c}{|\widehat{x}|^{\widehat{\Delta}}}\ . 
\end{align}
\paragraph{Sub-defect one-point function.} Clearly, the sub-defect one-point function $\langle\, \widecheck{\CO}_{\widecheck{\Delta}}(\check{\bm{x}})\,\rangle$ ($\check{\bm{x}}^{\mu}=(\check{x}^{a}, \hat{x}^{\alpha}=0, x_{\perp}^{i}=0)$) vanishes except the case where the sub-defect local operator is the identity one $\bm{1}$. We therefore have
\begin{align}
    \langle\, \widecheck{\CO}_{\widecheck{\Delta}}(\check{\bm{x}})\,\rangle=c\, \delta_{\, \widecheck{\CO}, \bm{1}}\ , 
\end{align}
where $c$ is again some constant.

We should note that conformal correlators with only defect and sub-defect local operators insertions are subject to the constraints of a conventional defect CFT whose bulk and defect are $p$ and $r$-dimensional flat spaces, respectively. Since those constraints have been well studied in \cite{Billo:2016cpy, Gadde:2016fbj}, we confine ourselves to the conformal correlators with at least one bulk local operator inserted below.

\paragraph{Bulk--sub-defect two-point function.} A first non-trivial example of two-point functions is the bulk--sub-defect two-point function:
\begin{align}
     \langle\, \CO_{\Delta}(X)\,\widecheck{\CO}_{\widecheck{\Delta}}(\widecheck{Y})\, \rangle\ .
\end{align}
By noting $\widecheck{Y}\bullet \widecheck{Y}=0$, there are three invariants:
\begin{align}
    X\circ X\qquad , \qquad X\diamond X\qquad , \qquad X\bullet \widecheck{Y}\ ,  
\end{align}
and only one cross-ratio $\xi$ defined in \eqref{eq: cross-ratio 1}. We therefore have the conformal correlator in the embedding space:
\begin{align}\label{eq: bulk-subdefect two pt}
    \langle\, \CO_{\Delta}(X)\,\widecheck{\CO}_{\widecheck{\Delta}}(\widecheck{Y})\,  \rangle=\frac{f(\xi)}{(X\circ X)^{(\Delta-\widecheck{\Delta})/2}\, (-2X\bullet \widecheck{Y})^{\widecheck{\Delta}}}\ , 
\end{align}
and in the physical space, it is reduced to
\begin{align}
    \langle\, \CO_{\Delta}(x)\,\widecheck{\CO}_{\widecheck{\Delta}}(\check{\bm{y}})\,  \rangle=\frac{f(\xi)}{|x_{\perp}|^{\Delta-\widecheck{\Delta}}\, |x-\check{\bm{y}}|^{2\widecheck{\Delta}}}\ .
\end{align}

\paragraph{Bulk--defect two-point function.} We next consider the bulk-defect two-point function $\langle\, \CO_{\Delta}(x)\,\widehat{\CO}_{\widehat{\Delta}}(\hat{\bm{y}})\, \rangle$ whose counterpart in the embedding space is 
\begin{align}
    \langle\, \CO_{\Delta}(X)\,\widehat{\CO}_{\widehat{\Delta}}(\widehat{Y})\, \rangle\ .
\end{align}
Since $X$ and $\widehat{Y}$ are of course subject to the null conditions, independent invariants constructed from them are given by
\begin{align}
    X\circ X\quad , \quad X\diamond X\quad , \quad \widehat{Y}\diamond \widehat{Y}\quad , \quad X\cdot \widehat{Y} \quad , \quad X\diamond \widehat{Y}\ . 
\end{align}
Likewise in the discussion of the bulk one-point function, we can also construct three cross-ratios:
\begin{align}
    \xi_{1}\equiv\frac{(X\circ X)^{1/2}\,(\widehat{Y}\diamond \widehat{Y})^{1/2}}{-2X\cdot \widehat{Y}}\quad , \quad \xi_{2}\equiv\frac{(X\circ X)^{1/2}\,(\widehat{Y}\diamond \widehat{Y})^{1/2}}{X\diamond\widehat{Y}}\quad , \quad \xi_{3}\equiv\xi \ ,
\end{align}
where $\xi$ is defined in \eqref{eq: cross-ratio 1}.
By imposing the homogeneity conditions, the bulk-defect two-point function is determined as
\begin{align}\label{eq: bulk-defect two pt}
    \langle\, \CO_{\Delta}(X)\,\widehat{\CO}_{\widehat{\Delta}}(\widehat{Y})\, \rangle=\frac{f(\xi_{1}, \xi_{2}, \xi_{3})}{(X\circ X)^{(\Delta-\widehat{\Delta})/2}\,(-2X\cdot \widehat{Y})^{\widehat{\Delta}}}\ . 
\end{align}
In the physical space, three cross-ratios and inner products are reduced to 
\begin{align}
    \xi_{1}\longrightarrow \frac{|x_{\perp}|\,|\hat{{y}}|}{|x-\hat{\bm{y}}|^{2}}\quad , \quad \xi_{2}\longrightarrow \frac{|x_{\perp}|\,|\hat{{y}}|}{\hat{x}\cdot \hat{y}}\quad , \quad 
    \xi_{3}\longrightarrow \frac{|x_{\perp}|^2}{|\hat{x}|^{2}} \quad  ,
\end{align}
\begin{align}
    X\circ X\longrightarrow |x_{\perp}|^{2}\qquad , \qquad -2 X\cdot \widehat{Y}\longrightarrow |x-\hat{\bm{y}}|^{2}\ , 
\end{align}
hence the physical space correlator is fixed in the following way:
\begin{align}
    \langle\, \CO_{\Delta}(x)\,\widehat{\CO}_{\widehat{\Delta}}(\hat{\bm{y}})\, \rangle=\frac{f(\xi_{1}, \xi_{2}, \xi_{3})}{|x_{\perp}|^{\Delta-\widehat{\Delta}}\, |x-\hat{\bm{y}}|^{2\widehat{\Delta}}}\ . 
\end{align}
We should notice that when $p-r=1$, the above three cross-ratios are not independent since $\xi_{2}$ can be written in terms of $\xi_{3}$. Therefore, in that case, the number of cross-ratios is reduced to two rather than three.  

We can also determine the form of correlators by using the method of images \cite{Nishioka:2022ook} which are discussed in appendix \ref{sec: method of images scalar}.

\subsection{Sub-defect operator expansions of bulk local scalar primary}\label{subsec: OE}
Operator product expansions (OPEs) have always played a crucial role in the analysis of CFTs even in the presence of a conformal defect. For instance, the OPE ensures that when two bulk local scalar primaries $\CO_{\Delta_{1}}(x_{1})$ and $\CO_{\Delta_{2}}(x_{2})$ are enough to close to each other, we can expand their product in terms of primaries and their descendants:
\begin{align}
    \CO_{\Delta_{1}}(x_{1})\times \CO_{\Delta_{2}}(x_{2})=\sum_{\Delta_{3}}\frac{c_{123}}{|x_{12}|^{\Delta_{1}+\Delta_{2}-\Delta_{3}}}\CO_{\Delta_{3}}(x_{2})+(\text{descendants})\ . 
\end{align}
The similar operator identities to the above OPE formula do hold for the product of two defect local operators $\widehat{\CO}_{\widehat{\Delta}_{1}}\times \widehat{\CO}_{\widehat{\Delta}_{2}}$ and two sub-defect ones $\widecheck{\CO}_{\widecheck{\Delta}_{1}}\times \widecheck{\CO}_{\widecheck{\Delta}_{2}}$. In addition. in defect CFTs, we can also expand a bulk local operator in terms of defect local ones. This expansion is often called \emph{defect operator expansion} (DOE). In section \ref{subsubsec: review of DOE}, we present a review of the notion of DOE in conventional defect CFTs. In section \ref{subsubsec: SDOE}, we apply the DOE to composite defect systems and show that a bulk local operator can also be expanded in terms of sub-defect local ones\footnote{In the context of wedge CFTs, the SDOE is discussed for concrete models \cite[equation (2.13)]{Bissi:2022bgu}, i.e., expanding a bulk local operator in terms of sub-defect ones in our language. We emphasize that our discussions here can be applied to any kind of composite defect CFTs.}. 

\subsubsection{Defect operator expansions (DOEs)}\label{subsubsec: review of DOE}
\begin{figure}[t]
    \centering
       \begin{tikzpicture}
        \coordinate (A) at (0 , 0) {};
        \coordinate (B) at (5 , 0) {};
        \coordinate (C) at (9 , 0) {};
        \coordinate (D) at (14 , 0) {};

        \fill[green!25] ($(A)!0.5!(B)$) circle (1.5);
        \draw[ultra thick, orange] (A) -- (B);
        \node[below=0.2cm, orange] at (B) {$\widehat{\CD}^{(p)}$};
        \node[above right=0.1cm] at ($(A)!0.5!(B)+(0, 1.5)$) {$\CO_{\Delta}(x)$};
        \filldraw[black,very thick] ($(A)!0.5!(B)+(0, 1.5)$) circle (0.1);
        \draw[thick, dotted] ($(A)!0.5!(B)+(0, 1.5)$) --($(A)!0.5!(B)$);
        \node at ($(A)!0.5!(B)$) {$\times$};
        \node[below right=0.06cm] at ($(A)!0.5!(B)$) {$\hat{\bm{x}}^{\mu}$};

        \node at ($(B)!0.5!(C)$) {{\huge $=$}};

        \fill[green!25] ($(C)!0.5!(D)$) circle (1.5);
        \draw[ultra thick, orange] (C) -- (D);
        \node[below=0.2cm, orange] at (D) {$\widehat{\CD}^{(p)}$};
        \node[above right=0.1cm] at ($(C)!0.5!(D)$) {$\widehat{\CO}_{\widehat{\Delta}}(\hat{\bm{x}})$};
        \filldraw[black,very thick] ($(C)!0.5!(D)$) circle (0.1);
         \node at ($(B)!0.5!(C)+(1.2, 0)$) {\Large $\sum_{\widehat{\Delta}}$};
     \end{tikzpicture} 
    \caption{Schematic picture of the DOE. A bulk local operator $\CO_{\Delta}(x)$ can be expanded by using the defect local ones $\widehat{\CO}_{\widehat{\Delta}}(\hat{\bm{x}})$. This expansion is valid only when no local operators are inserted inside of the virtual ball (green region) whose center is $\hat{\bm{x}}^{\mu}$ and radius is $|x_{\perp}|$.}
    \label{fig: picture of doe}
\end{figure}
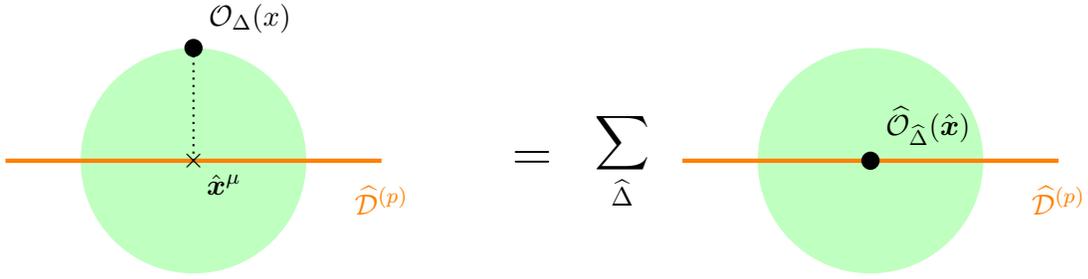
To explain the concept of the DOE, let us consider a conventional defect CFT with a single $p$-dimensional conformal defect. We can expand the bulk local scalar primary $\CO_{\Delta}(x)$ in terms of defect one $\widehat{\CO}_{\widehat{\Delta}}(\hat{\bm{x}})$ as follows:
\begin{align}\label{eq: doe}
    \CO_{\Delta}(x)=\sum_{\widehat{\Delta}}\frac{b_{\Delta, \widehat{\Delta}}}{|x_{\perp}|^{\Delta-\widehat{\Delta}}}\, \CB^{(d, p)}_{\widehat{\Delta}}\left(|x_{\perp}|, \widehat{\bm{\Box}}_{p}\right)\widehat{\CO}_{\widehat{\Delta}}(\hat{\bm{x}})\ + \ \text{(spinning contributions)}\ ,
\end{align}
where $\widehat{\bm{\Box}}_{p}$ is the $p$-dimensional Laplacian associated to the conformal defect, and $\CB^{(d, p)}_{\widehat{\Delta}}\left(|x_{\perp}|, \widehat{\bm{\Box}}_{p}\right)$ is a differential operator which is given by \cite[equation (B.3)]{Billo:2016cpy}
\begin{align}\label{eq: def of B}
    \CB^{(d, p)}_{\widehat{\Delta}}\left(|x_{\perp}|, \widehat{\bm{\Box}}\right)=\sum_{n=0}^{\infty}\frac{(-4)^{-n}}{n!\, \left(\widehat{\Delta}-\frac{p-2}{2}\right)_{n}}\, \ |x_{\perp}|^{2n}\, \widehat{\bm{\Box}}_{p}^{n}\ .  
\end{align}
Also, coefficients $\{b_{\Delta, \widehat{\Delta}}\}$ appearing in the DOE are often called DOE coefficients which cannot be fixed only from the symmetry and characterize the CFT data of the conformal defect. We should comment here that the DOE formula of $\CO_{\Delta}(x)$ does hold only when no other local operators are inserted inside the virtual ball centered at $\hat{\bm{x}}^{\mu}$ with radius $|x_{\perp}|$ (see figure \ref{fig: picture of doe}.). This remark plays a crucial role in the next section.

\subsubsection{Sub-defect operator expansions (SDOEs)}\label{subsubsec: SDOE}
In the previous section, we reviewed the DOE which is a peculiar operator identity to defect CFTs. By utilizing the DOE, we can also expand a bulk local operator in terms of sub-defect local operators in composite defect CFTs. We refer to this expansion as \emph{sub-defect operator expansion} (SDOE) throughout this paper. The main aim of this section is to derive the SDOE formula paying our attention to only sub-defect scalar channels. 

Our strategy for deriving the SDOE formula for a bulk local primary $\CO_{\Delta}(x)$ consists of the following two steps (see figure \ref{fig: picture of SDOE}):
\begin{itemize}
    \item \textbf{Step 1.} we expand a bulk local operator $\CO_{\Delta}(x)$ located in $|x_{\perp}|<|\hat{x}|$ in terms of defect local ones $\{\widehat{\CO}_{\widehat{\Delta}}(\hat{\bm{x}})\}$.
    \item \textbf{Step 2.} we further expand each defect local operator appearing in \textbf{Step 1} in terms of sub-defect ones $\{\widecheck{\CO}_{\widecheck{\Delta}}(\check{\bm{x}})\}$. 
\end{itemize}
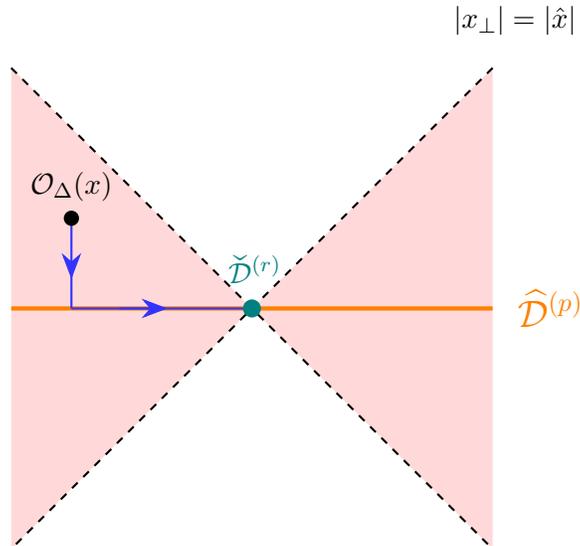
\begin{figure}[t]
    \centering
       \begin{tikzpicture}[scale=0.8]
        \coordinate (A) at (0 , 0) {};
        \coordinate (B) at (8 , 0) {};
        \fill[red!15] ($(A)!0.5!(B)+(4, 4)$)--($(A)!0.5!(B)$)--($(A)!0.5!(B)+(4, -4)$);
        \fill[red!15] ($(A)!0.5!(B)+(-4, 4)$)--($(A)!0.5!(B)$)--($(A)!0.5!(B)+(-4, -4)$);
        \draw[ultra thick, orange] (A) -- (B);
        \draw[thick, dashed] ($(A)!0.5!(B)+(4, 4)$) --($(A)!0.5!(B)$)--($(A)!0.5!(B)+(-4, -4)$);
        \draw[thick, dashed] ($(A)!0.5!(B)+(-4, 4)$) --($(A)!0.5!(B)$)--($(A)!0.5!(B)+(4, -4)$);
        \node[teal, above=0.2cm] at ($(A)!0.5!(B)+(0.05,0)$) {$\widecheck{\CD}^{(r)}$};
        \draw [arrows = {-Stealth[scale=1.5]}, blue!80, thick] ($(A)!0.5!(B)+(-3, 1.5)$) -- ($(A)!0.5!(B)+(-3, 1.5)!0.675!(-3, 0)$);
        \draw[blue!80, thick] ($(A)!0.5!(B)+(-3, 0.75)$)-- ($(A)!0.5!(B)+(-3, 0)$);
        \draw [arrows = {-Stealth[scale=1.5]}, blue!80, thick] ($(A)!0.5!(B)+(-3, 0)$) -- ($(A)!0.325!(B)$);
        \draw[blue!80, thick] ($(A)!0.25!(B)-(0.2,0)$)-- ($(A)!0.5!(B)$);
        \node[right=0.2cm, orange] at (B) {{\Large $\widehat{\CD}^{(p)}$}};
        \node[above=0.1cm] at ($(A)!0.5!(B)+(-3, 1.5)$) {$\CO_{\Delta}(x)$};
        \filldraw[black,very thick] ($(A)!0.5!(B)+(-3, 1.5)$) circle (0.1);
        \node[above right=1cm] at (6.3, 3.5) {$|x_{\perp}|=|\hat{x}|$};
        \fill[teal] ($(A)!0.5!(B)$) circle (0.15);
     \end{tikzpicture} 
    \caption{Our strategy for deriving the SDOE of a bulk local operator $\CO_{\Delta}(x)$. We first expand $\CO_{\Delta}(x)$ in terms of defect local operators (\textbf{Step 1}), and next expand defect local operators by using sub-defect operators (\textbf{Step 2}). To define the SDOE correctly, we need to place a bulk local operator $\CO_{\Delta}(x)$ at a bulk point such that $|x_{\perp}|<|\hat{x}|$ (depicted as the red region). See also figure \ref{fig: picture of SDOE region}.}
    \label{fig: picture of SDOE}
\end{figure}
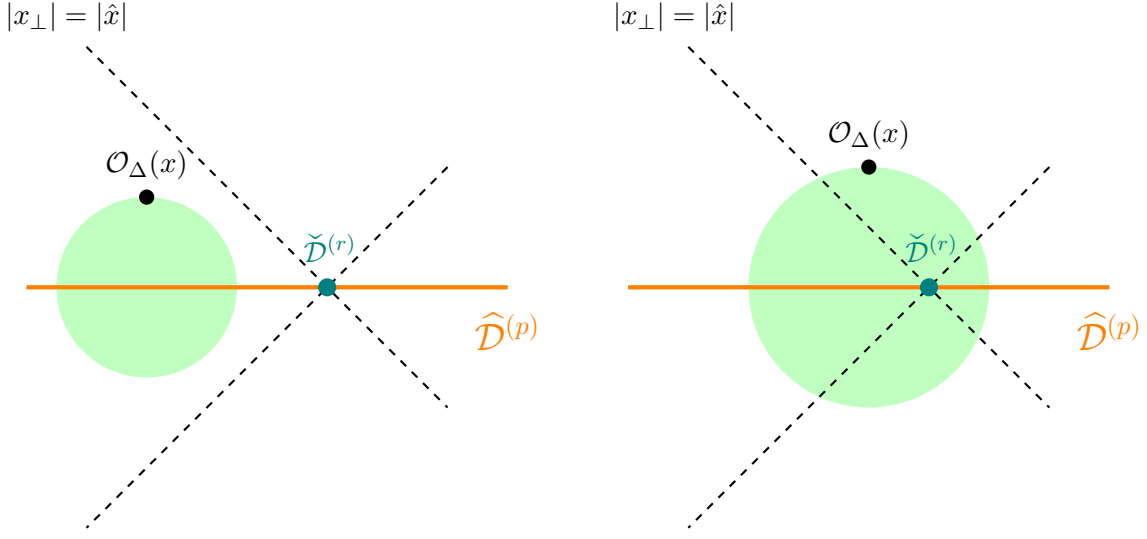
\begin{figure}[t]
    \centering
    \begin{tikzpicture}[scale=0.8]
        \coordinate (A) at (0 , 0) {};
        \coordinate (B) at (8 , 0) {};
        \fill[green!25] ($(A)!0.5!(B)+(-3, 0)$) circle (1.5);
        \draw[ultra thick, orange]  ($(A)-(1,0)$)--($(B)-(1,0)$);
        \draw[thick, dashed] ($(A)!0.5!(B)+(2, 2)$) --($(A)!0.5!(B)$)--($(A)!0.5!(B)+(-4, -4)$);
        \draw[thick, dashed] ($(A)!0.5!(B)+(-4, 4)$) --($(A)!0.5!(B)$)--($(A)!0.5!(B)+(2, -2)$);
        \node[teal, above=0.2cm] at ($(A)!0.5!(B)+(0.05,0)$) {$\widecheck{\CD}^{(r)}$};
        \node[below=0.2cm, orange] at ($(B)-(1,0)$) {{\Large $\widehat{\CD}^{(p)}$}};
        \node[above=0.1cm] at ($(A)!0.5!(B)+(-3, 1.5)$) {$\CO_{\Delta}(x)$};
        \filldraw[black,very thick] ($(A)!0.5!(B)+(-3, 1.5)$) circle (0.1);
        \node[above right=1cm] at (-2.4, 3.2) {$|x_{\perp}|=|\hat{x}|$};
        \fill[teal] ($(A)!0.5!(B)$) circle (0.15);

        \coordinate (C) at (10 , 0) {};
        \coordinate (D) at (18 , 0) {};
        \fill[green!25] ($(C)!0.5!(D)+(-1, 0)$) circle (2);
        \draw[ultra thick, orange] ($(C)-(1,0)$)--($(D)-(1,0)$);
        \draw[thick, dashed] ($(C)!0.5!(D)+(2, 2)$) --($(C)!0.5!(D)$)--($(C)!0.5!(D)+(-4, -4)$);
        \draw[thick, dashed] ($(C)!0.5!(D)+(-4, 4)$) --($(C)!0.5!(D)$)--($(C)!0.5!(D)+(2, -2)$);
        \node[teal, above=0.2cm] at ($(C)!0.5!(D)+(0.05,0)$) {$\widecheck{\CD}^{(r)}$};
        \node[below=0.2cm, orange] at ($(D)-(1,0)$) {{\Large $\widehat{\CD}^{(p)}$}};
        \node[above=0.1cm] at ($(C)!0.5!(D)+(-1, 2)$) {$\CO_{\Delta}(x)$};
        \filldraw[black,very thick] ($(C)!0.5!(D)+(-1, 2)$) circle (0.1);
        \node[above right=1cm] at (7.7, 3.2) {$|x_{\perp}|=|\hat{x}|$};
        \fill[teal] ($(C)!0.5!(D)$) circle (0.15);
     \end{tikzpicture} 
    \caption{In order to implement our strategy for defining the SDOE, we need to place a bulk local operator $\CO_{\Delta}(x)$ such that $|x_{\perp}|<|\hat{x}|$ as depicted in the left panel. If we do so, we can take the DOE of $\CO_{\Delta}(x)$ without any issues since the sub-defect is out of the virtual ball (green region). On the one hand, if we put $\CO_{\Delta}(x)$ such that $|x_{\perp}|>|\hat{x}|$ like in the right panel, we can no longer define the DOE because the sub-defect does exist inside of the virtual ball.}
    \label{fig: picture of SDOE region}
\end{figure}
In \textbf{Step 1}, we carry out the DOE for a bulk local field $\CO_{\Delta}(x)$ based on \eqref{eq: doe}. Importantly, for the well-definedness of this DOE, we need to put $\CO_{\Delta}(x)$ at a bulk point such that $|x_{\perp}|<|\hat{x}|$. If we do so, indeed, we are sure to find a virtual ball in which the sub-defect does not exist as depicted in the left panel of figure \ref{fig: picture of SDOE region}. This implies that we can always expand a bulk local operator by the complete set of defect operators. On the other hand, however, if we place $\CO_{\Delta}(x)$ such that $|x_{\perp}|<|\hat{x}|$, the sub-defect does always enter inside of the virtual ball as depicted in the right panel of figure \ref{fig: picture of SDOE region}. This means that we cannot expand a bulk local operator only in terms of defect ones, and we must take into account the effect of the sub-defect. For these reasons, whenever we consider the SDOE of a bulk local operator, we put a bulk local operator such that $|x_{\perp}|<|\hat{x}|$.

To implement \textbf{Step 2}, it is sufficient to make use of the DOE formula again:
\begin{align}\label{eq: subdoe}
    \widehat{\CO}_{\widehat{\Delta}}(\widehat{\bm{x}})=\sum_{\widecheck{\Delta}}\frac{b_{\widehat{\Delta}, \widecheck{\Delta}}}{|\hat{x}|^{\widehat{\Delta}-\widecheck{\Delta}}}\, \CB^{(p, r)}_{\widecheck{\Delta}}\left(|\hat{x}|, \widecheck{\bm{\Box}}_{r}\right)\widecheck{\CO}_{\widecheck{\Delta}}(\check{\bm{x}})\ + \ \text{(spinning contributions)}\ , 
\end{align}
where $\widecheck{\bm{\Box}}_{r}$ is the $r$-dimensional Laplacian associated to the sub-defect $\widecheck{\CD}^{(r)}$. To avoid some tediousness, we particularly focus on sub-defect scalar channels in this paper. We also leave the generalization to the sub-defect spinning ones for interesting future work. By combining the above two operator expansions \eqref{eq: doe} and \eqref{eq: subdoe}, we can expand a bulk local operator in terms of sub-defect local operators:
\begin{align}
    \CO_{\Delta}(x)=\sum_{\widecheck{\Delta}}\sum_{\widehat{\Delta}}\frac{b_{\Delta, \widehat{\Delta}}\, b_{\widehat{\Delta}, \widecheck{\Delta}}}{|x_{\perp}|^{\Delta-\widehat{\Delta}}}\,\CB^{(d, p)}_{\widehat{\Delta}}\left(|x_{\perp}|, \widehat{\bm{\Box}}_{p}\right)\frac{1}{|\hat{x}|^{\widehat{\Delta}-\widecheck{\Delta}}}\CB^{(p, r)}_{\widecheck{\Delta}}\left(|\hat{x}|, \widecheck{\bm{\Box}}_{r}\right)\widecheck{\CO}_{\widecheck{\Delta}}(\check{\bm{x}})\ .
\end{align}

By using \eqref{eq: def of B} and somewhat non-trivial calculations, we eventually arrive at the following SDOE formula:
\begin{align}\label{eq: SDOE formula}
    \begin{aligned}
    &\CO_{\Delta}(x) \\
    &\  = \sum_{\widecheck{\Delta}}\sum_{\widehat{\Delta}}\frac{b_{\Delta, \widehat{\Delta}}\, b_{\widehat{\Delta}, \widecheck{\Delta}}}{|x_{\perp}|^{\Delta-\widecheck{\Delta}}}\sum_{L=0}^{\infty}\sum_{m=0}^{L}\, \frac{1}{(L-m)!\, m!\, \left(\widehat{\Delta}-\frac{p-2}{2}\right)_{L-m}\left(\widecheck{\Delta}-\frac{r-2}{2}\right)_{m} }\,\xi^{-m+\frac{\widehat{\Delta}-\widecheck{\Delta}}{2}}\, \\
    &\ \ \cdot{}_2 F_1\left(-m+\frac{\widehat{\Delta}-\widecheck{\Delta}}{2}, 1-m+\frac{-p+r+\widehat{\Delta}-\widecheck{\Delta}}{2};1+L-m-\frac{p}{2}+\widehat{\Delta};-\xi \right)\left(-\frac{|\hat{x}|^{2}}{4}\,\widecheck{\bm{\Box}}_{\,r}\right)^{L}\,\widecheck{\CO}_{\widecheck{\Delta}}(\check{\bm{x}})\ , 
    \end{aligned}
\end{align}
where  $\xi$ is the cross-ratio which is defined in \eqref{eq: cross-ratio 1}, and takes its value in the range of $0<\xi<1$. Also, ${}_2 F_{1}$ is the hypergeometric function defined by
\begin{align}
    {}_2 F_{1}(a, b;c;z)\equiv \sum_{n=0}^{\infty}\frac{(a)_n\, (b)_n}{(c)_n n!}z^{n}\qquad , \qquad |z|<1\ .  
\end{align}
In the above expression, we should remark that while the index $L=0$ corresponds to the sub-defect primary channel $\widecheck{\CO}_{\widecheck{\Delta}}$, the index $L\geq 1$ does to the level of the sub-defect descendant channel $\widecheck{\bm{\Box}}^{\,L}_{\,r}\,\widecheck{\CO}_{\widecheck{\Delta}}$. To the best of our efforts, deriving the SDOE formula presented above into a closed form turns out to be quite challenging in general. Nonetheless, we will illustrate some examples where the SDOEs can be expressed in closed forms in section \ref{subsec: O(N) free scalar model}.

\subsection{Conformal block expansions}\label{subsec: conformal blocks}
In section \ref{subsec: correlator}, we derive various conformal correlators in composite defect CFTs. In this section, we particularly focus on a bulk one-point function \eqref{eq: bulk one pt} and a bulk--sub-defect two-point one \eqref{eq: bulk-subdefect two pt}. These correlators are not completely fixed by the residual conformal symmetry, and they admit the conformal block expansions. We dedicate this section to derive the conformal block expansions for these correlators.
\subsubsection{Bulk one-point function} 
We begin with exploring the conformal expansion of a bulk one-point function $\langle\,\CO_{\Delta}(x) \, \rangle$. We now take the origin of the radial quantization at $x$, draw a $(d-1)$-dimensional sphere tangent to the defect $\widehat{\CD}^{(p)}$, and insert a completeness relation:
\begin{align}\label{eq: completeness condition}
    \begin{aligned}
    \bm{1}&=\sum_{\widehat{\Delta}} |\, \widehat{\CO}_{\widehat{\Delta}}\, |\\
    &\equiv\sum_{\widehat{\Delta}}\left(\ket{\widehat{\CO}_{\widehat{\Delta}}}\bra{\widehat{\CO}_{\widehat{\Delta}}}+(\text{descendants})\right)\ . 
    \end{aligned}
\end{align}
Here, the state $\ket{\widehat{\CO}_{\widehat{\Delta}}}$ is the diagonal basis of the SO$(1, p+1)\times\text{SO}(d-p)$ Casimir operator $\widehat{Cas}$:\footnote{Remark that the defect local primaries are classified by the representations of SO$(1, p+1)\times\text{SO}(d-p)$ rather than SO$(1, r+1)\times\text{SO}(d-p)\times\text{SO}(p-r)$}
\begin{align}
    \widehat{Cas}\, \ket{\widehat{\CO}_{\widehat{\Delta}}}= \widehat{\Delta}(\widehat{\Delta}-p)\, \ket{\widehat{\CO}_{\widehat{\Delta}}} \quad , \quad \widehat{Cas}\equiv -\frac{1}{2}\bJ_{AB}\bJ^{AB}-\frac{1}{2}\bJ_{ij}\bJ^{ij}\ , 
\end{align}  
where $A, B=-1, 0, \cdots , p$ and $i, j=p+1, p+2, \cdots , d$. Since the Casimir operator $ \widehat{Cas}$ particularly commutes with momentum operators along with $\widehat{\CD}^{(p)}$, the following relation does hold:
\begin{align}\label{eq: casimir relation}
    \widehat{Cas}\, |\, \widehat{\CO}_{\widehat{\Delta}}\, |=|\, \widehat{\CO}_{\widehat{\Delta}}\, |\,  \widehat{Cas}=\widehat{\Delta}(\widehat{\Delta}-p)\, |\, \widehat{\CO}_{\widehat{\Delta}}\, |\ . 
\end{align}
In the discussion below, we consider the embedding space correlator $\langle\,\CO_{\Delta}(X) \, \rangle$ instead of the physical space one $\langle\,\CO_{\Delta}(x) \, \rangle$ since the Casimir operator in the embedding space acts linearly on the operator. Under these preparations, we can expand a bulk one-point function $\langle\,\CO_{\Delta}(X) \, \rangle$ in the following way:
\begin{align}\label{eq: expansion}
    \begin{aligned}
    \langle\,\CO_{\Delta}(X) \, \rangle&=\sum_{\widehat{\Delta}}\bra{\CD^{(p, r)}}\, \widehat{\CO}_{\widehat{\Delta}}\, |\, \CO_{\Delta}(X) \ket{0} \\ 
    &\equiv\sum_{\widehat{\Delta}}\frac{c_{\widehat{\Delta}}}{(X\circ X)^{\Delta/2}}\, G_{\widehat{\Delta}}(\xi)\ , 
    \end{aligned}
\end{align}
where $\ket{\CD^{(p, r)}}$ and  $\ket{0}$ are the conformal vacuum states which are invariant under the actions of SO$(1, r+1)\times\text{SO}(d-p)\times\text{SO}(p-r)$ and SO$(1, d+1)$, respectively. Also, $c_{\widehat{\Delta}}$ is some expansion coefficient, and $G_{\widehat{\Delta}}(\xi)$ is the what is called \emph{conformal block}. In what follows, we derive the exact form of the conformal block $G_{\widehat{\Delta}}(\xi)$ and the relation between $c_{\widehat{\Delta}}$ and OE coefficients.
By using the relation \eqref{eq: casimir relation}, we can derive a non-trivial differential equation as follows;
\begin{align}\label{eq: diff1}
    \begin{aligned}
        \widehat{\Delta}(\widehat{\Delta}-p)\, \bra{\CD^{(p, r)}}\, \widehat{\CO}_{\widehat{\Delta}}\, |\, \CO_{\Delta}(X) \ket{0} & =\bra{\CD^{(p, r)}}\, \widehat{\CO}_{\widehat{\Delta}}\, |\,\widehat{Cas}\, \CO_{\Delta}(X) \ket{0}\\
        &=-\frac{1}{2}\CL_{AB}\CL^{AB}(X)\,\bra{\CD^{(p, r)}}\, \widehat{\CO}_{\widehat{\Delta}}\, |\, \CO_{\Delta}(X) \ket{0}\ , 
    \end{aligned}
\end{align}  
where $\CL_{MN}(X)$ is the differential operator which is defined by
\begin{align}
    \CL_{MN}(X)\equiv X_{M}\frac{\partial}{\partial X^{N}}-X_{N}\frac{\partial}{\partial X^{M}}\ . 
\end{align} 
Notice that both $X\circ X$ and $\xi$ are SO($d-p$) invariant, hence we do not need to take into account the differential operator $\CL_{ij}\CL^{ij}(X)$. After somewhat tedious computations, the above equation \eqref{eq: diff1} is reduced to 
\begin{align}\label{eq: casimir eq bulk one pt}
    4\xi^{2}\,(\xi+1)\frac{\d^{2}}{\d\xi^{2}}\, G_{\widehat{\Delta}}(\xi)-2\xi\left(p-2+(p-r-4)\xi \right)\frac{\d}{\d\xi}\, G_{\widehat{\Delta}}(\xi)-\widehat{\Delta}(\widehat{\Delta}-p)G_{\widehat{\Delta}}(\xi)=0\ . 
\end{align} 
To solve this differential equation, we need to specify the boundary condition for the conformal block $G_{\widehat{\Delta}}$ at $\xi=0$. With the help from the DOE \eqref{eq: doe}, the bulk one-point function should show the following asymptotic behavior:
\begin{align}
    \langle\,\CO_{\Delta}(X) \, \rangle\quad \overset{{\footnotesize \xi\to 0}}{\longrightarrow}\quad \sum_{\widehat{\Delta}}\frac{b_{\Delta, \widehat{\Delta}}}{(X\circ X)^{(\Delta-\widehat{\Delta})/2}}\,\langle\,\widehat{\CO}_{\widehat{\Delta}}(\widehat{X}) \, \rangle=\sum_{\widehat{\Delta}}\frac{b_{\Delta, \widehat{\Delta}}\, b_{\widehat{\Delta}, 0}}{(X\circ X)^{\Delta/2}}\, \xi^{\widehat{\Delta}/2} \ , 
\end{align}
from which we can read off the expansion coefficient $c_{\widehat{\Delta}}$ in \eqref{eq: expansion}, and the boundary condition for $G_{\widehat{\Delta}}$:
\begin{align}\label{eq: boundary condition}
    c_{\widehat{\Delta}}=b_{\Delta, \widehat{\Delta}}\, b_{\widehat{\Delta}, 0}\qquad , \qquad G_{\widehat{\Delta}}(\xi)\ \overset{{\footnotesize \xi\to 0}}{\longrightarrow}\ \xi^{\widehat{\Delta}/2}\ . 
\end{align}
Under this boundary condition, the solution for the differential equation \eqref{eq: casimir eq bulk one pt} can be computed to be\footnote{Since the differential equation \eqref{eq: casimir eq bulk one pt} is invariant under $\widehat{\Delta}\to p-\widehat{\Delta}$, there is another independent solution which can be obtained by replacing $\widehat{\Delta}$ with $p-\widehat{\Delta}$ in \eqref{eq: conformal block bulk one}. This solution corresponds to the conformal block of the defect shadow operator, and we exclude this solution by imposing the boundary condition \eqref{eq: boundary condition}.}
\begin{align}\label{eq: conformal block bulk one}
    G_{\widehat{\Delta}}(\xi)=\xi^{\widehat{\Delta}/2}\, {}_2 F_1\left(\frac{\widehat{\Delta}}{2}, \frac{2-p+r+\widehat{\Delta}}{2}\,;\, \frac{2-p}{2}+\widehat{\Delta}\, ;\, -\xi \right)\ . 
\end{align}
In summary, the conformal block expansion of a bulk one-point function in the physical space is given by
\begin{align}
    \langle\,\CO_{\Delta}(x) \, \rangle=\sum_{\widehat{\Delta}}\frac{b_{\Delta, \widehat{\Delta}}\, b_{\widehat{\Delta}, 0}}{|x_{\perp}|^{\Delta}}\, G_{\widehat{\Delta}}(\xi)\ ,
\end{align}
where the conformal block $G_{\widehat{\Delta}}(\xi)$ is given by \eqref{eq: conformal block bulk one}. This expansion is completely consistent with the SDOE formula \eqref{eq: SDOE formula}.
\subsubsection{Bulk--sub-defect two-point function}
We next consider the conformal block expansion of a bulk--sub-defect two-point function $\langle\, \CO_{\Delta}(x)\,\widecheck{\CO}_{\widecheck{\Delta}}(\check{\bm{y}})\,  \rangle$. Since our methodology is the same as the case of a bulk one-point function, we just present the result. A bulk--sub-defect two point function admit the following conformal block expansion: 
\begin{align}
    \langle\, \CO_{\Delta}(X)\,\widecheck{\CO}_{\widecheck{\Delta}}(\widecheck{Y})\,  \rangle=\sum_{\widehat{\Delta}}\frac{b_{\Delta, \widehat{\Delta}}\, b_{\widehat{\Delta}, \widecheck{\Delta}}}{|x_{\perp}|^{\Delta-\widecheck{\Delta}}\, |x-\check{\bm{y}}|^{2\widecheck{\Delta}}}\, \widetilde{G}_{\widehat{\Delta}}(\xi)\ , 
\end{align}  
where the conformal block $\widetilde{G}_{\widehat{\Delta}}(\xi)$ is given by
\begin{align}
    \widetilde{G}_{\widehat{\Delta}}(\xi)=\xi^{(\widehat{\Delta}-\widecheck{\Delta})/2}\,{}_2 F_1\left(\frac{\widehat{\Delta}-\widecheck{\Delta}}{2}, \frac{2-p+r+\widehat{\Delta}-\widecheck{\Delta}}{2}\, ;\, \frac{2-p}{2}+\widehat{\Delta}\,;\,-\xi \right)\ . 
\end{align}
As a consistency check, if we set the sub-defect primary to be identity operator, namely $\widecheck{\Delta}=0$, the above expansion is reduced to \eqref{eq: conformal block bulk one}.

\section{O$(N)$ free scalar model in composite defect CFT}\label{subsec: O(N) free scalar model}
Although we have seen general aspects of composite defect CFTs so far, the discussions were somewhat abstract. We therefore use this section to provide instructive examples of composite defect CFTs by paying attention to the O$(N)$ free scalar model. For simplicity, we only consider the fundamental O$(N)$ scalar field $\phi^{I}$ $(I=1, 2, \cdots , N)$ which is subject to Neumann or Dirichlet boundary conditions, and we assume that the O($N$) global symmetry is manifestly preserved even on the composite defect. 

Firstly, the bulk two-point function can be derived by solving the following Poisson equation:
\begin{align}
    \bm{\Box}_{d, x}\, \langle\, \phi^{I}(x)\, \phi^{J}(y)\, \rangle=\frac{4\pi^{d/2}}{\Gamma(d/2)}\, \delta^{IJ}\, \delta^{(d)}(x-y)\ , 
\end{align}
in the presence of the composite defect $\CD^{(p. r)}$. By using the method of image, we can solve this as follows:
\begin{align}\label{eq: bulk twopt O(N) scalar}
    \langle\, \phi^{I}(x)\, \phi^{J}(y)\, \rangle
    =\delta^{IJ}\left[\frac{1}{|x-y|^{d-2}}+\frac{A}{|x-y'|^{d-2}}+\frac{B}{|x-\overline{y}|^{d-2}}+\frac{C}{|x-\overline{y}'|^{d-2}}\right]\ , 
\end{align}
where $A, B$ and $C$ are some numerical constants that depend on the boundary condition of the fundamental scalar $\phi^{\, I}$ and $y'$, $\overline{y}$ and $\overline{y}'$ are mirror configurations (see also \eqref{eq: mirror config}):
\begin{align}
    y'\equiv(\check{y}^{a}, \hat{y}^{\alpha}, -y^{i})\qquad , \qquad \overline{y}\equiv(\check{y}^{a}, -\hat{y}^{\alpha}, -y^{i})\qquad , \qquad \overline{y}'\equiv(\check{y}^{a}, -\hat{y}^{\alpha}, y^{i})\ . 
\end{align}
We now have two kinds of conformal defects $\widehat{\CD}^{(p)}$ and $\widecheck{\CD}^{(r)}$, hence we can accordingly consider the following four types of boundary conditions (BCs) of a bulk local operator $\phi^{\, I}$ in principle:
\begin{itemize}
    \item (N, N)-type\, :\, Neumann for $\widehat{\CD}^{(p)}$ and Neumann for $\widecheck{\CD}^{(r)}$\ , 
    \item (N, D)-type\, :\, Neumann for $\widehat{\CD}^{(p)}$ and Dirichlet for $\widecheck{\CD}^{(r)}$\ , 
    \item (D, N)-type\, :\, Dirichlet for $\widehat{\CD}^{(p)}$ and Neumann for $\widecheck{\CD}^{(r)}$\ , 
    \item (D, D)-type\, :\, Dirichlet for $\widehat{\CD}^{(p)}$ and Dirichlet for $\widecheck{\CD}^{(r)}$\ , 
\end{itemize}
where ``Neumann for $\widehat{\CD}^{(p)}$'' and ``Dirichlet for $\widehat{\CD}^{(p)}$'' are defined by
\begin{align}
    \text{Neumann for $\widehat{\CD}^{(p)}$}\ &: \ \left.\partial_{i}\, \phi^{\, I}\right|_{x_{\perp j}=0\, , \, \hat{x}_{\beta}\not=0}=0\ , \\ 
    \text{Dirichlet for $\widehat{\CD}^{(p)}$}\ &: \ \left. \phi^{\, I}\right|_{x_{\perp j}=0\, , \, \hat{x}_{\beta}\not=0}=0\ . 
\end{align}
Also, ``Neumann for $\widecheck{\CD}^{(r)}$'' and ``Dirichlet for $\widecheck{\CD}^{(r)}$'' are denoted by
\begin{align}
    \text{Neumann for $\widecheck{\CD}^{(r)}$}\ &: \ \left.\partial_{i}\, \phi^{\, I}\right|_{x_{\perp j}=0\, , \, \hat{x}_{\beta}=0}=\left.\partial_{\alpha}\, \phi^{\, I}\right|_{x_{\perp j}=0\, , \, \hat{x}_{\beta}=0}=0\ , \\
    \text{Dirichlet for $\widecheck{\CD}^{(r)}$}\ &: \ \left. \phi^{\, I}\right|_{x_{\perp j}=0\, , \, \hat{x}_{\beta}=0}=0\ .
\end{align}
\begin{table}[t]
    \centering
    \begin{tabular}{c|ccc}
              & $A$ & $B$ & $ C$  \\ \hline
       (N, N) &  $1$  & $1$   & $1$     \\
       (N, D) &  $1$  & $-1$   & $-1$    \\
       (D, N) &  $-1$  & $1$   & $-1$     
    \end{tabular}
    \caption{Numerical constants appearing in the bulk two-point function \eqref{eq: bulk twopt O(N) scalar}. Each BC of the fundamental scalar $\phi^{\, I}$ corresponds to the three-tuple $(A, B, C)$.   
    }
    \label{table: numerical constants}
\end{table}
We can, however, convince ourselves that the (D, D)-type BC is reduced to the conventional defect CFT since the sub-defect $\widecheck{\CD}^{(r)}$ is embedded into $\widehat{\CD}^{(p)}$. We therefore do not pursue the (D, D)-type BC case further. For the other three types of BCs, we can fix the numerical constants $A, B$ and $C$ after simple computations. The results can be seen in table \ref{table: numerical constants}. For simplicity, we set the defect and sub-defect dimensions such that $p=d-1$ and $r=d-2$ except for section \ref{subsec: em tensor} and section \ref{subsubsec: ANEC constraint}.

\subsection{Defect CFT data}\label{subsec: defect CFT data}
Although it is a quite hard problem to fully understand defect and sub-defect operator spectrums in general, we can provide strong constraints on these spectrums when a bulk theory is free. In our model, we have the fundamental O$(N)$ bulk field $\phi^{I}$ which satisfies the following equation of motion (EOM):
\begin{align}\label{eq: eom}
    \bm{\Box}_{\, d}\, \phi^{I}(x)=0\ , 
\end{align}
where $ \bm{\Box}_{\, d}$ is the $d$-dimensional Laplacian. In this section, we draw fully upon the strength of this EOM to restrict defect operator spectrums. We moreover derive the explicit form of DOEs of $\phi^{I}(x)$ for both Neumann and Dirichlet BCs. 

\subsubsection{Defect operator spectrums}\label{subsec: defect operator spectrum}
Suppose that the DOE of the bulk field $\phi^{I}$ contains a defect primary $\widehat{\CO}^{\,I}_{\widehat{\Delta}}$ whose conformal dimension is $\widehat{\Delta}$ (see \eqref{eq: doe}):
\begin{align}
    \phi^{I}(x)\, \supset\,b_{\phi, \widehat{\CO}}\ x_{\perp}^{-\frac{d-2}{2}+\widehat{\Delta}}\ \widehat{\CO}^{\,I}_{\widehat{\Delta}}(\hat{\bm{x}})\ , 
\end{align}
where $b_{\phi, \widehat{\CO}}\,(\not=0)$ is some OE constant between $\phi^{I}$ and $\widehat{\CO}^{\,I}_{\widehat{\Delta}}$. We also use the symbol $\supset$ to single out only defect primary sector and drop the descendant contributions. The EOM \eqref{eq: eom} then implies the conformal dimensions $\widehat{\Delta}$ must satisfy the following constraint:
\begin{align}
    \left(\widehat{\Delta}-\frac{d-2}{2}\right)\left(\widehat{\Delta}-\frac{d}{2}\right)=0\quad \Longleftrightarrow \quad \widehat{\Delta}=\frac{d-2}{2} \quad \text{or} \quad \frac{d}{2}\ .
\end{align} 
While $\hat{\phi}^{\, I}\equiv \widehat{\CO}^{\,I}_{(d-2)/2}$ is the defect primary in the case of Neumann BC, and $\partial_{\perp}\, \hat{\phi}^{\, I}\equiv \widehat{\CO}^{\,I}_{d/2}$ is the one in the case of Dirichlet one. We should note that these defect primaries can be deduced from the bulk field $\phi^{I}$ by taking the following limits:
\begin{align}\label{eq: defect primaries}
    \begin{aligned}
        &\text{Neumann for } \widehat{\CD}^{(d-1)}\ : \ \hat{\phi}^{\,I}(\hat{\bm{x}})\equiv \lim_{x_{\perp}\to0}\phi^{I}(x) \ ,\\
        &\text{Dirichlet for } \widehat{\CD}^{(d-1)}\ : \ \partial_{\perp}\hat{\phi}^{\,I}(\hat{\bm{x}})\equiv \lim_{x_{\perp}\to0}\, x_{\perp}^{-1}\,\phi^{I}(x)\ . 
    \end{aligned}
\end{align}

\subsubsection{Defect operator expansion}
In the previous section, we have succeeded in identifying the full defect operator spectrums. In this subsection, we derive the exact form of the DOE of a bulk field $\phi^{I}$ by utilizing the result of the previous section. Let us start from the DOE in the case of the Neumann BC for the defect $\widehat{\CD}^{(d-1)}$. 

\paragraph{Neumann BC for $\widehat{\CD}^{(d-1)}$.} To get the DOE , it is sufficient to compute the following bulk-defect and defect-defect two-point functions:
\begin{align}
    \langle\, \phi^{I}(x)\, \hat{\phi}^{\,J}(\hat{\bm{y}})\, \rangle\qquad , \qquad \langle\, \hat{\phi}^{\,I}(\hat{\bm{x}})\, \hat{\phi}^{\,J}(\hat{\bm{y}})\, \rangle\ , 
\end{align}
where the hatted field $\hat{\phi}^{\,I}$ is the defect local primary which is defined in \eqref{eq: defect primaries}. Both of them can be deduced from the bulk two-point function \eqref{eq: bulk twopt O(N) scalar} by taking the limits $y^{d}_{\perp}\to 0$ and $(x^{d}_{\perp}, y^{d}_{\perp})\to 0\,$, respectively:
\begin{align}
        \langle\, \phi^{I}(x)\, \hat{\phi}^{\,J}(\hat{\bm{y}})\, \rangle &= \frac{\delta^{IJ}}{|x-\hat{\bm{y}}|^{d-2}}\left[A+1+(B+C)\left({\eta+1}\right)^{-\frac{d-2}{2}}\right]\ ,   \label{eq: bulk_defect func Neumann defect}\\
        \langle\, \hat{\phi}^{\,I}(\hat{\bm{x}})\, \hat{\phi}^{\,J}(\hat{\bm{y}})\, \rangle &= \frac{\delta^{IJ}}{|\hat{\bm{x}}-\hat{\bm{y}}|^{d-2}}\left[A+1+(B+C)\left(\hat{\eta}+1\right)^{-\frac{d-2}{2}}\right]\ ,  \label{eq: defect_defect func Neumann defect}
\end{align}
where $\eta$ and $\hat{\eta}$ are cross-ratios which are defined by
\begin{align}\label{eq: cross-ratio 2}
    \eta\equiv \frac{4\, \hat{x}\cdot\hat{y}}{|x-\hat{\bm{y}}|^2}\qquad , \qquad \hat{\eta}\equiv \frac{4\, \hat{x}\cdot\hat{y}}{|\hat{\bm{x}}-\hat{\bm{y}}|^2}\ . 
\end{align}
As a consistency check, we should notice that when the BC is (D, N)-type, both of the above two-point functions vanish. These results \eqref{eq: bulk_defect func Neumann defect} and \eqref{eq: defect_defect func Neumann defect} imply that the DOE of $\phi^{I}$ which is subject to the Neumann BC should be as follows:
\begin{align}\label{eq: OE for Neumann wrt defect}
    \phi^{I}(x)=\sum_{n=0}^{\infty}\frac{(-4)^{-n}}{n!\, \left(\frac{1}{2}\right)_{n}}\, \ x_{\perp}^{2n}\, \widehat{\bm{\Box}}_{d-1}^{n}\,\hat{\phi}^{\,I}(\hat{\bm{x}}) \  .
\end{align}
The proof is given in appendix \ref{appendix: proof of OEs}. We should notice that this operator expansion is the same as \eqref{eq: doe} and not affected by the presence of the sub-defect $\widecheck{\CD}^{(d-2)}$ even though the proof requires somewhat non-trivial computations. This is, however, not surprising since the operator expansion is an operator identity that is defined \emph{locally} as emphasized in section \ref{subsec: OE}.   
\paragraph{Dirichlet BC for $\widehat{\CD}^{(d-1)}$.} In a similar manner to the above case, we can derive the DOE of a bulk field $\phi^{\, I}$ which is subject to the Dirichlet BC for $\widehat{\CD}^{(d-1)}$. In this case, the defect local primary that we should consider is $\partial_{\perp}\hat{\phi}^{\, I}$ which has the conformal dimension $d/2$ (see \eqref{eq: defect primaries} for its definition.). In the same way as \eqref{eq: bulk_defect func Neumann defect} and \eqref{eq: defect_defect func Neumann defect}, we can obtain the bulk-defect and defect-defect two-point functions concerning $\partial_{\perp}\hat{\phi}$ from \eqref{eq: bulk twopt O(N) scalar}:
\begin{align}
    \langle\, \phi^{I}(x)\, \partial_{\perp}\hat{\phi}^{\,J}(\hat{\bm{y}})\, \rangle &= \frac{(d-2)\, \delta^{IJ}}{|x-\hat{\bm{y}}|^{d}}\left[-A+1+(C-B)\left({\eta+1}\right)^{-\frac{d}{2}}\right]x_{\perp}\ ,   \label{eq: bulk_defect func Dirichlet defect}\\
    \langle\, \partial_{\perp}\hat{\phi}^{\,I}(\hat{\bm{x}})\, \partial_{\perp}\hat{\phi}^{\,J}(\hat{\bm{y}})\, \rangle &= \frac{(d-2)\,\delta^{IJ}}{|\hat{\bm{x}}-\hat{\bm{y}}|^{d}}\left[-A+1+(C-B)\left(\hat{\eta}+1\right)^{-\frac{d}{2}}\right]\ ,  \label{eq: defect_defect func Dirichlet defect}
\end{align}
where $\eta$ and $\hat{\eta}$ are cross-ratios defined in \eqref{eq: cross-ratio 2}. After the similar computations performed in appendix \ref{appendix: proof of OEs}, we can read off the DOE of a bulk field $\phi^{\,I}$ which is subject to the Dirichlet BC:
\begin{align}\label{eq: DOE Dirichlet}
    \phi^{I}(x)=\sum_{n=0}^{\infty}\frac{(-4)^{-n}}{n!\, \left(\frac{3}{2}\right)_{n}}\, \ x_{\perp}^{2n+1}\, \widehat{\bm{\Box}}_{d-1}^{n}\,\partial_{\perp}\hat{\phi}^{\,I}(\hat{\bm{x}}) \  .
\end{align}
We can observe again that this DOE formula is completely irrelevant to the sub-defect $\widecheck{\CD}^{(d-2)}$ as expected.
\subsection{Sub-defect CFT data}\label{subsec: subdefect data}
In section \ref{subsec: defect CFT data}, we identified the defect operator spectrums and derived the DOEs of the fundamental bulk field $\phi^{I}$ which is subject to Neumann or Dirichlet BCs. In this section, we perform a similar analysis to the previous section and explore the sub-defect CFT data.
\subsubsection{Sub-defect operator spectrums}\label{eq: subdefect op spect}
 In this section, we specify the sub-defect operator spectrums by employing the same method as section \ref{subsec: defect operator spectrum}. 
Assuming  that the SDOE of the bulk field $\phi^{\,I}$ has a contribution from a sub-defect primary $\widecheck{\CO}_{\widecheck{\Delta}}(x)$ whose conformal dimension is given $\widecheck{\Delta}$ (see \eqref{eq: SDOE formula}):
\begin{align}\label{eq: SDOE O(N) vector model Neumann}
    \phi^{I}(x)\, \supset\, \sum_{\widehat{\Delta}}\frac{b_{\phi, \widehat{\Delta}}\, b_{\widehat{\Delta}, \widecheck{\Delta}}}{x_{\perp}^{\, d/2-1-\widecheck{\Delta}}}\,\xi^{\frac{\widehat{\Delta}-\widecheck{\Delta}}{2}}\,{}_2 F_1\left(\frac{\widehat{\Delta}-\widecheck{\Delta}}{2}, \frac{1+\widehat{\Delta}-\widecheck{\Delta}}{2};\frac{3-d}{2}+\widehat{\Delta};-\xi \right)\widecheck{\CO}_{\widecheck{\Delta}}(\check{\bm{x}})\ . 
\end{align}
In what follows, we utilize the EOM \eqref{eq: eom} to this SDOE formula and investigate the sub-defect operator spectrums. 

We first consider the case where a bulk field $\phi^{\, I}(x)$ is subject to the Neumann BC for $\widehat{\CD}^{(d-1)}$. As discussed in section \ref{subsec: defect operator spectrum}, we have the unique defect local primary $\hat{\phi}^{\, I}(\hat{\bm{x}})$ which is defined in \eqref{eq: defect primaries}, and the above summation over all defect primaries is reduced to the single contribution. 
Keeping this in mind, acting the $d$-dimensional Laplacian $ \bm{\Box}_{d}$ on both hand sides of \eqref{eq: SDOE O(N) vector model Neumann} leads to 
\begin{align}\label{eq: inter eom}
    \begin{aligned}
    \bm{\Box}_{d}\,\phi^{I}(x)\, \supset\, \frac{b_{\hat{\phi}\, , \, \widecheck{\Delta}}}{x_{\perp}^{d/2+1-\widecheck{\Delta}}}\,\xi^{\frac{d-2-2\widecheck{\Delta}}{4}}\, \frac{d-2-2\widecheck{\Delta}}{48}\, H(\widecheck{\Delta}, \xi)\, \widecheck{\CO}_{\widecheck{\Delta}}(\check{\bm{x}})\ , 
    \end{aligned}
\end{align}
where $H(\widecheck{\Delta}, \xi)$ is given by:
\begin{align}
    \begin{aligned}
        &H(\widecheck{\Delta}, \xi)\\
        &\, \equiv (d-2\widecheck{\Delta})(d-2\widecheck{\Delta}+2)(d-2\widecheck{\Delta}+4)\xi^{3/2}(1+\xi)\, {}_2 F_1\left(\frac{6+d-2\widecheck{\Delta}}{4}, \frac{8+d-2\widecheck{\Delta}}{4}\,;\,\frac{5}{2}\, ;\, -\xi \right) \\
        &\quad +12(1+\xi)^{\frac{-d+2\widecheck{\Delta}}{4}}\left[(d-2\widecheck{\Delta})\, \xi^{1/2}(1+\xi)^{1/2}\cos\left(\frac{d-2\widecheck{\Delta}-2}{2}\tan^{-1}(\sqrt{\xi})\right)\right. \\
        &\qquad\qquad\qquad\qquad\qquad\qquad\qquad\qquad\left. -2\left(1+(d-2\widecheck{\Delta}+1)\xi\right)\, \sin\left(\frac{d-2\widecheck{\Delta}}{2}\tan^{-1}(\sqrt{\xi})\right) \right]\ . 
    \end{aligned}
\end{align}
In order for the EOM to hold, the right-hand side of \eqref{eq: inter eom} needs to identically vanish regardless of the location of a bulk local field. We can observe that this vanishment condition is established only when $\widecheck{\Delta}=(d-2)/2$ or $\widecheck{\Delta}=d/2$. We therefore conclude that $\check{\phi}^{\, I}\equiv \widecheck{\CO}^{\,I}_{(d-2)/2}$ and $\widehat{\partial}\, \check{\phi}^{\, I}\equiv \widecheck{\CO}^{\,I}_{d/2}$ correspond to the sub-defect primaries in the case of the Neumann and Dirichlet BCs for the sub-defect $\widecheck{\CD}^{(d-2)}$, respectively. In a similar manner to \eqref{eq: defect primaries}, we should note that these sub-defect primaries can be deduced from the defect primary $\hat{\phi}^{\,I}$ by taking the following limits:
\begin{align}\label{eq: subdefect primaries Neumann}
    \begin{aligned}
        &\text{(N, N)}\quad : \quad \check{\phi}^{\,I}(\check{\bm{x}})\equiv \lim_{\hat{x}\to0}\hat{\phi}^{\,I}(\hat{\bm{x}}) \ ,\\
        &\text{(N, D)}\quad : \quad \widehat{\partial}\,\check{\phi}^{\,I}(\check{\bm{x}})\equiv \lim_{\hat{x}\to0}\, \hat{x}^{-1}\,\hat{\phi}^{\,I}(\hat{\bm{x}})\ . 
    \end{aligned}
\end{align}  

We finally describe the case where the bulk field $\phi^{\, I}$ obeys the Dirichlet BC for the defect $\widehat{\CD}^{(d-1)}$. After similar computations to the above, it turns out that the only sub-defect primaries with $\widecheck{\Delta}=d/2$ or $\widecheck{\Delta}=(d+2)/2$ do contribute to the SDOE. As stated in the discussion below the equation \eqref{eq: bulk twopt O(N) scalar}, we do not consider the (D, D)-type BC, hence we drop the sub-defect primary with $\widecheck{\Delta}=d/2$. On the one hand, $\widehat{\partial}\,\partial_{\perp}\, \check{\phi}^{\, I}\equiv \widecheck{\CO}^{\,I}_{(d+2)/2}$ is the unique sub-defect primary for the (D, N)-type BC which can also be deduced from the defect primary $\partial_{\perp}\hat{\phi}^{\,I}$ by the following limit:
\begin{align}
    \text{(D, N)}\quad : \quad \widehat{\partial}\,\partial_{\perp}\,\check{\phi}^{\,I}(\check{\bm{x}})\equiv \lim_{\hat{x}\to0}\, \hat{x}^{-1}\,\partial_{\perp}\,\hat{\phi}^{\,I}(\hat{\bm{x}})\ . 
\end{align}
\begin{table}[t]
    \centering
\renewcommand{\arraystretch}{1.7}
\begin{tabular}{M{2cm}|M{3cm}|M{3cm}} \hline
    & Defect ops. & Sub-defect ops. \\\hline
(N, N)& \multirow{2}{*}{$\hat{\phi}^{\, I}$}&$\check{\phi}^{\, I}$ \\ \cline{1-1} \cline{3-3}
(N, D)&                          &$\widehat{\partial}\,\check{\phi}^{\, I}$ \\ \hline
(D, N)&  $\partial_{\perp}\, \hat{\phi}^{\, I}$  & $\widehat{\partial}\,\partial_{\perp}\,\check{\phi}^{\, I}$ \\ \hline
\end{tabular}
\caption{Defect and sub-defect operator spectrums in the O$(N)$ free vector model.   
}
\label{table: operator spectrums}
\end{table}
In table \ref{table: operator spectrums}, we summarize all defect and sub-defect operator contents of the O$(N)$ free model.

\subsubsection{Sub-defect operator expansion}\label{subsubsec: SDOE O(N)}
As described in section \ref{subsec: OE}, we can also expand the field $\phi^{I}$ in terms of the sub-defect operators. The main aim of this section is to obtain the SDOE formulas of a bulk field $\phi^{I}$. As we discussed at the beginning of section \ref{subsec: O(N) free scalar model}, we have three kinds of BCs: (N, N), (N, D), and (D, N)-types. We will see the SDOE for each case in order. 
\paragraph{Case 1: (N, N)\,.} We first derive the SDOE when the bulk field $\phi^{I}$ is subject to Neumann BCs for both conformal defects $\widehat{\CD}^{(d-1)}$ and $\widecheck{\CD}^{(d-2)}$. The simplest way to get this formula is to consider the SDOE of the defect operator $\hat{\phi}$\,:
\begin{align}
    \hat{\phi}^{I}(\hat{\bm{x}})=\sum_{n=0}^{\infty}\frac{(-4)^{-n}}{(n!)^{2}}\, \ \hat{x}^{2n}\, \widecheck{\bm{\Box}}_{d-2}^{\,n}\,\check{\phi}^{\,I}(\check{\bm{x}}) \  ,
\end{align}
which can be also obtained from \eqref{eq: bulk twopt O(N) scalar} in a similar way to DOEs. Plugging this expression into \eqref{eq: OE for Neumann wrt defect} leads to
\begin{align}\label{eq: SDOE NN type 1}
    \phi^{I}(x)=\sum_{n, m=0}^{\infty}\sum_{k=0}^{n}\frac{(-4)^{-n-m}\,\Gamma(2m+1)}{(m!)^2\, k!\, (n-k)!\, (1/2)_{n}\, \Gamma(2m-2k+1)}x_{\perp}^{2n}\hat{x}^{2m-2k} \, \widecheck{\bm{\Box}}_{d-2}^{\,n-k+m}\,\check{\phi}^{\,I}(\check{\bm{x}})\ . 
\end{align}
To evaluate this horrible summation, it is convenient to change the sum variables as follows:
\begin{align}
    \sum_{n, m=0}^{\infty}\sum_{k=0}^{n}\, f(n, m, k)=\sum_{L=0}^{\infty}\,\sum_{n=0}^{L}\,\sum_{m=L-n}^{L}\, f(n, m, n+m-L)\ ,  
\end{align}
where $f$ is an arbitrary function that satisfies 
\begin{align}\label{eq: condition f}
    f(n, m, k)=0\qquad  \quad \text{for}\qquad m<k\ . 
\end{align}
By applying this rearrangement formula to \eqref{eq: SDOE NN type 1}, the SDOE becomes\footnote{We should notice that the summand in \eqref{eq: SDOE NN type 1} satisfies \eqref{eq: condition f} thanks to the factor $\Gamma(2m-2k+1)$ appearing in the denominator.}
\begin{align}
    \phi^{I}(x)=\sum_{L=0}^{\infty}\,\sum_{n=0}^{L}\,\sum_{m=L-n}^{L}\frac{(-4)^{-n-m}\,(2m)!}{(m!)^2\, (n+m-L)!\, (L-m)!\, (1/2)_{n}\, (2L-2n)!}\, \xi^{n}\,(\hat{x}^{2} \, \widecheck{\bm{\Box}}_{d-2})^{\,L}\,\check{\phi}^{\,I}(\check{\bm{x}})\ . 
\end{align}
We moreover employ the following summation identities:
\begin{align}
    \begin{aligned}
        \sum_{m=L-n}^{L}\frac{(-4)^{-m}\, (2m)!}{(m!)^{2}\, (n+m-L)!\, (L-m)!}&=\frac{(-1)^{L+n}\, \Gamma(1/2+L-n)\,\Gamma(1/2+n)}{\pi\, L!\, n!}\ , \\
        \sum_{n=0}^{L}\frac{1}{(L-n)!\, n!}\,\xi^{n}&=\frac{(1+\xi)^{L}}{L!}\ , 
    \end{aligned}
\end{align}
and we obtain a quite simple form of the SDOE as follows:
\begin{align}\label{eq: SDOE formula N, N case}
    \phi^{I}(x)=J_{0}\left((\hat{x}^2 +x_{\perp}^2)\, \widecheck{\bm{\Box}}_{d-2}\right)\, \check{\phi}^{\,I}(\check{\bm{x}})\ ,
\end{align}
where $J_{0}(\chi)$ is the Bessel function of the first kind:\footnote{We remark that in this definition of the Bessel function, we employ unconventional notation compared with the mathematical literature.}
\begin{align}
    J_{0}(\chi)\equiv\sum_{m=0}^{\infty}\frac{1}{4^{m}\, (m!)^2}\,(-\chi)^{m}\ . 
\end{align}
We should comment here that the above SDOE formula \eqref{eq: SDOE formula N, N case} is completely consistent with the following bulk to sub-defect two-point function:
\begin{align}
    \langle\, \phi^{I}(x)\, \check{\phi}^{\,J}(\check{\bm{y}})\, \rangle &= \frac{4\,\delta^{IJ}}{|x-\check{\bm{y}}|^{d-2}}\ ,
\end{align} 
which can be again deduced from \eqref{eq: bulk twopt O(N) scalar}.
\paragraph{Case 2: (N, D)\,.} We next consider the SDOE of the bulk field $\phi^{I}$ which obeys the Neumann BC for the defect $\widehat{\CD}^{(d-1)}$ and the Dirichlet one for the sub-defect $\widecheck{\CD}^{(d-2)}$. Our strategy closely follows the previous case. First, we can expand the defect primary operator $\hat{\phi}^{\, I}$ in terms of the sub-defect primary $\widehat{\partial}\,\check{\phi}^{\,I}$ and its descendants:
\begin{align}
\hat{\phi}^{I}(\hat{\bm{x}})=\sum_{n=0}^{\infty}\frac{(-4)^{-n}}{n!\, \left(2\right)_{n}}\, \ \hat{x}^{2n+1}\, \widecheck{\bm{\Box}}_{d-2}^{n}\,\widehat{\partial}\,\check{\phi}^{\,I}(\check{\bm{x}}) \  .
\end{align}
After substituting this expression into \eqref{eq: OE for Neumann wrt defect} and performing a similar computation to the previous case, we arrive at the following SDOE formula:
\begin{align}
    \phi^{I}(x)=\hat{x}\cdot\, {}_{0}F_{1}\left(2\, ; \, -\frac{\hat{x}^2+x_{\perp}^{2}}{4}\, \widecheck{\bm{\Box}}_{d-2}\right)\widehat{\partial}\,\check{\phi}^{I}(\bm{x})\ , 
\end{align}
where ${}_{0}F_{1}(a\,;\,z)$ is a confluent hypergeometric function which is defined as
\begin{align}
    {}_{0}F_{1}(a\,;\,z)\equiv \sum_{k=0}^{\infty}\frac{\Gamma(a)}{\Gamma(a+k)\, k!}\,z^{k}\ . 
\end{align}
\paragraph{Case 3: (D, N)\,.} We next move on to the final case, namely Dirichlet BC for the defect $\widehat{\CD}^{(d-1)}$ and the Neumann one for the sub-defect $\widecheck{\CD}^{(d-2)}$. In this case, there is a unique non-trivial defect primary $\partial_{\perp}\hat{\phi}^{I}$, and its SDOE consists of the sub-defect primary $\widehat{\partial}\,\partial_{\perp}\check{\phi}^{\,I}$ and its descendants:
\begin{align}
    \partial_{\perp}\hat{\phi}^{\,I}(\hat{\bm{x}})=\sum_{n=0}^{\infty}\frac{(-4)^{-n}}{n!\, (3)_{n}}\, \hat{x}^{2n+1}\, \widecheck{\bm{\Box}}_{d-2}^{\,n}\, \widehat{\partial}\,\partial_{\perp}\check{\phi}^{\,I}(\check{\bm{x}})\ . 
\end{align}
By plugging this into \eqref{eq: DOE Dirichlet} and similar computations to the above two cases, we can obtain the SDOE formula of a bulk field $\phi^{\,I}$:
\begin{align}
    \phi^{\, I}(x)=\hat{x}\,x_{\perp}\cdot\, {}_{0}F_{1}\left(3\, ; \, -\frac{\hat{x}^2+x_{\perp}^{2}}{4}\, \widecheck{\bm{\Box}}_{d-2}\right)\widehat{\partial}\,\partial_{\perp}\,\check{\phi}^{\,I}(\check{\bm{x}})\ . 
\end{align}
Even in this case, we can readily verify that this SDOE formula is consistent with the following two-point functions:
\begin{align}
    \langle\, \phi^{I}(x)\, \widehat{\partial}\,\partial_{\perp}\,\check{\phi}^{\,J}(\check{\bm{y}})\, \rangle &= \frac{4\,d\,(d-2)\, \delta^{IJ}}{|x-\check{\bm{y}}|^{d+2}}x_{\perp}\, \hat{x}\ ,   \label{eq: bulk_sdefect func DN}\\
    \langle\, \widehat{\partial}\,\partial_{\perp}\,\check{\phi}^{\,I}(\check{\bm{x}})\, \widehat{\partial}\,\partial_{\perp}\,\check{\phi}^{\,J}(\check{\bm{y}})\, \rangle &= \frac{4\,d\,(d-2)\, \delta^{IJ}}{|x-\check{\bm{y}}|^{d+2}}\ .  \label{eq: sdefect_sdefect func DN}
\end{align}

\subsection{Energy-momentum tensor in the O($N$) vector model}\label{subsec: em tensor}
One of the key ingredients in CFT is the energy-momentum (EM) tensor, hence we here particularly derive the bulk one-point function of the EM tensor $T_{\mu\nu}$ which is given by\footnote{Unlike the previous sections, we do not limit the dimensions of a defect and a sub-defect in this section and the next section.}
\begin{align}
    T_{\mu\nu}=\partial_{\mu}\phi^{I}\partial_{\nu}\phi_{I}-\frac{1}{2}\delta_{\mu\nu}\partial^{\rho}\phi^{I}\partial_{\rho}\phi_{I}-\frac{d-2}{4(d-1)}(\partial_{\mu}\partial_{\nu}-\delta_{\mu\nu}\, \bm{\Box}_{d})\,|\phi|^2\ . 
\end{align}
To compute this quantity, we first need to derive the bulk one-point function of $|\phi|^2$. This can be done by setting $y\to x$ in \eqref{eq: bulk twopt O(N) scalar} and subtracting the divergence which is inconsistent with the conformal symmetry:
\begin{align}
   \langle\, |\phi|^{2}(x) \, \rangle=\frac{N}{2^{d-2}}\frac{G(\xi)}{|x_{\perp}|^{d-2}} \ , 
\end{align}
where $G(\xi)$ is the function defined by
\begin{align}
    G(\xi)\equiv A + B\xi^{\frac{d-2}{2}}(1+\xi)^{-\frac{d-2}{2}}++ C\, \xi^{\frac{d-2}{2}}\ . 
\end{align}
After tedious calculations, we can compute the bulk one-point function of the EM tensor:
\begin{align}\label{eq: energy_momentum tensor one-pt}
    \begin{aligned}
        \langle\, T_{ab}\, \rangle&= \frac{F_{d}}{|x_{\perp}|^{d}}\left[(d-p-1)\, A +(d-r-1)\, B\, \xi^{\frac{d}{2}}(1+\xi)^{-\frac{d}{2}}+(p-r-1)\, C\, \xi^{\frac{d}{2}}\right]\, \delta_{ab} \ , \\
        \langle\, T_{\alpha\beta}\, \rangle&= \frac{F_{d}}{|x_{\perp}|^{d}}\left\{\left[(d-p-1)\, A -(r+1)\, B\, \xi^{\frac{d}{2}}(1+\xi)^{-\frac{d}{2}}-(d-p+r+1)\, C\, \xi^{\frac{d}{2}}\right]\, \delta_{\alpha\beta}\right.\\ &\left. \qquad\qquad\qquad\qquad +d\, \xi^{\frac{d}{2}+1}\left(B\, (1+\xi)^{-\frac{d}{2}-1}+C\right)\frac{\hat{x}_{\alpha}\, \hat{x}_{\beta}}{|x_{\perp}|^2}  \right\}\ , \\
        \langle\, T_{ij}\, \rangle&= \frac{F_{d}}{|x_{\perp}|^{d}}\left\{\left[-(1+p)\, A -(r+1)\, B\, \xi^{\frac{d}{2}}(1+\xi)^{-\frac{d}{2}}+(p-r-1)\, C\, \xi^{\frac{d}{2}}\right]\, \delta_{ij}\right.\\ &\left. \qquad\qquad\qquad\qquad +d\, \left(A+B\,  \xi^{\frac{d}{2}+1}(1+\xi)^{-\frac{d}{2}-1}\right)\frac{x_{\perp i}\, x_{\perp j}}{|x_{\perp}|^2}   \right\}\ , \\
        \langle\, T_{\alpha i}\, \rangle&=\frac{F_{d}}{|x_{\perp}|^{d}}\, d\,B\, \xi^{\frac{d}{2}+1}(1+\xi)^{-\frac{d}{2}-1}\,\frac{\hat{x}_{\alpha}\, x_{\perp i}}{|x_{\perp}|^2}\ , \\
        \langle\, T_{a \alpha}\, \rangle&=\langle\, T_{a i}\, \rangle=0\ , 
    \end{aligned}
\end{align}
where $F_{d}$ is the constant which is defined by
\begin{align}
    F_{d}\equiv\frac{N(d-2)}{2^{d}(d-1)}\ . 
\end{align}
We should notice that the above expression is consistent with the conservation law $\partial_{\mu}T^{\mu\nu}=0$ and the traceless condition $T^{\mu}_{\ \mu}=0$. 

\subsection{Constraints from averaged null energy condition}\label{subsubsec: ANEC constraint}
In this section, we derive some constraints that come from the averaged null energy condition (ANEC). The ANEC asserts that the integral of the tangential-tangential component of the EM tensor over the light-ray trajectory must be semi-positive:
\begin{align}\label{eq: ANEC}
    \text{ANEC : }\quad \mathcal{E}\equiv\int_{-\infty}^{+\infty}\, \d \lambda\, \bra{\psi}\, T(\lambda) \, \ket{\psi}\geq 0\quad , \quad T(\lambda)\equiv \frac{\d\gamma^{\mu}}{\d\lambda}\frac{\d\gamma^{\nu}}{\d\lambda} T_{\mu\nu}(\gamma(\lambda)) \ , 
\end{align}
where $\gamma^{\mu}(\lambda)$ is a null geodesic whose trajectory is parametrized by the affine parameter $\lambda$, and $\ket{\psi}$ is an arbitrary state. Although this condition can be proven for a spacetime without any conformal defects \cite{Faulkner:2016mzt,Hartman:2016lgu,Kravchuk:2018htv}, it has not yet been proved for a spacetime in the presence of a conformal defect. Following \cite{Jensen:2018rxu,Chalabi:2021jud, Herzog:2020bqw}, nevertheless, we here assume that the ANEC does hold even for the defect systems, and investigate some constraints resulting from the ANEC below. From the physical reasoning, we also assume that the sub-defect $\widecheck{\CD}^{(r)}$ contains the time coordinate $t\equiv \i \check{x}^{1}$, and the conformal symmetry-breaking pattern in the Lorentzian signature should be rewritten as
    \begin{align}\label{eq: residual symmetry in Lorentzian signature}
        \text{SO}(2, d)\longrightarrow \text{SO}(2, r)\times \text{SO}(p-r)\times \text{SO}(d-p)\ . 
    \end{align}
A null geodesic coordinates $\gamma^{\mu}$ which is far away from the origin by a distance $R$ satisfy
\begin{align}\label{eq: geodesic coordi}
    -\d t^2+\sum_{a=2}^{r}\d\check{x}_{a}^{2}+\sum_{\alpha=r+1}^{p}\d\hat{x}_{\alpha}^{2} +\sum_{i=p+1}^{d}\,\d x_{\perp i }^2 =0\ \ , \ \ - t^2+\sum_{a=2}^{r}\d\check{x}_{a}^{2}+\sum_{\alpha=2}^{p}\hat{x}_{\alpha}^2 +\sum_{i=p+1}^{d}\, (x_{\perp}^i)^2 =R^{2}\ .
\end{align}
We should note that the residual conformal symmetry \eqref{eq: residual symmetry in Lorentzian signature} makes this set-up much simpler. Indeed, by using a rotational symmetry $\SO(r)\times\SO(p-r)\times\SO(d-p-1)$, we can move to the following coordinate:
\begin{align}
    (t\, , \, \check{x}^{2}\, , \,  0\, ,\cdots\, , 0\, , \, \hat{x}^{r+1} \, , \,  0\, ,\cdots\, , 0\, , \, x_{\perp}^{p+1}\, , \, x_{\perp}^{p+2}\, , \,  0\, ,\cdots\, , 0)\ , 
\end{align}
and we omit the zero coordinates below.
We furthermore can make use of the scaling symmetry and set $R=1$ in \eqref{eq: geodesic coordi}. Under these preparations, we can parametrize the null geodesic coordinates by an affine parameter $\lambda\in\BR$ with fixed two angles $\psi\in[0, \pi/2]$ and $\theta\in [0, 2\pi)$ as follows:
\begin{align}
    t=\lambda\quad , \quad \check{x}^{2}=\lambda\cos\psi \quad , \quad \hat{x}^{r+1}=\lambda\sin\psi\cos\theta  \quad , \quad x_{\perp}^{p+1}=\lambda\sin\psi\sin\theta \quad , \quad x_{\perp}^{p+2}=1\ . 
\end{align} 
By analytically continuing the bulk one-point function \eqref{eq: energy_momentum tensor one-pt} in Euclidean signature to the one in Lorentzian one, we can obtain the bulk one-point function of the null energy $\langle\, T(\lambda)\, \rangle$ as follows:
\begin{align}
    \langle\, T(\lambda)\, \rangle=-d\, F_{d}\left[\frac{A\, \sin^{2}\psi \sin^{2}\theta}{(1+\lambda^{2}\sin^{2}\psi \sin^{2}\theta)^{d/2+1}}+\frac{B\,\sin^{2}\psi}{(1+\lambda^{2}\sin^{2}\psi)^{d/2+1}}\right]\ . 
\end{align}
Interestingly, we observe that the dependence of defect and sub-defect dimensions has disappeared  in this expression. By plugging this into the ANEC \eqref{eq: ANEC} and using the following integral formula:
\begin{align}
    \int_{-\infty}^{\infty}\,\d\lambda\, \frac{1}{(1+b\, \lambda^{2})^{c}}=\frac{\Gamma(c-1/2)}{\Gamma(c)}\sqrt{\frac{\pi}{b}}\qquad , \qquad b\, , c>0 \ ,  
\end{align}
we can obtain the non-trivial constraints on the coefficients $A$ and $B$:
\begin{align}
    \mathcal{E}=-\frac{N\,\sqrt{\pi}\,\Gamma\left(\frac{d-1}{2}\right)\, \sin\psi\, (|\sin\theta|A+B)}{2^{d-1} \,  \Gamma\left(\frac{d-2}{2}\right)}\geq 0\ \ , \ \ \forall\psi\in[0, \pi/2]\ , \ \forall\theta\in[0, 2\pi) 
\end{align}  
which is equivalently 
\begin{align}\label{eq: ANEC constraint p=2, r=1}
    |\sin\theta|\, A +B \leq 0\qquad , \qquad  \forall\theta\in[0, 2\pi) 
\end{align}
We should notice that any constraints on the coefficient $C$ do not emerge from the ANEC. Quite surprisingly, by combining this with the BCs discussed in section \ref{subsec: O(N) free scalar model} (see table \ref{table: numerical constants}), it turns out that only (N, D)-type BC survives while (N, N) and (D, N) ones are excluded from the viewpoint of the ANEC constraint. This result seems to be somewhat curious since particularly, the composite defect CFT with (N, N)-type BC clearly satisfies the reflection positivity (recall that the ANEC is closely related to the unitarity.). We leave this puzzle for an open question.

\section{Conclusion and future directions}\label{sec: conclusion and future}
In this paper, we have investigated various aspects of the composite defect CFTs. In section \ref{sec: kinematical constraint}, we discussed general properties which do hold for any composite defect CFTs. Specifically, we determined the forms of the correlation functions, and contemplated the sub-defect operator expansion of a bulk local operator. We also derived the conformal block expansions for a bulk one-point and bulk--sub-defect two-point functions. Moreover, we demonstrated how these general results can work by using a free O$(N)$ vector model which is the simplest example of composite defect CFTs. In addition to this, we discussed the constraints from the ANEC. There are a lot of future directions concerning to this work, and some of which are listed in order. 
\begin{itemize}
    \item \textbf{Construction of composite defect: }we can create a new conformal defect by fusing two ones whose co-dimensions are the same as each other \cite{Soderberg:2021kne, SoderbergRousu:2023zyj, Diatlyk:2024zkk}. It remains an intriguing question to construct our composite defect $\CD^{(p, r)}$ by the fusion of two different defects $\widehat{\CD}^{(p)}$ and $\widecheck{\CD}^{(r)}$. 
    \item \textbf{Conformal dynamics of other models: }it is interesting to explore the dynamics of composite defect CFTs for other models. Particularly, this paper treated the free O$(N)$ model where a bulk O$(N)$ symmetry is manifestly preserved even on the composite defect. We can, in principle, consider the composite defect CFTs for interacting theories including not only scalars but also fermions and gauge fields. It may be also intriguing to introduce supersymmetries. For instance, in the four-dimensional $\CN=4$ super Yang-Mills theory, we can utilize the integrable property to extract the non-trivial CFT data \cite{Cavaglia:2021bnz}. By employing the integrability to our composite defect systems, it might be possible to constrain the sub-defect CFT data for supersymmetric systems. Finally, we can also consider the possibility that a bulk global symmetry is partially or completely broken on a sub-defect $\widecheck{\CD}^{(r)}$, which helps us understand the phase structures of composite defect CFTs.
    \item \textbf{Holographic interpretation: }we can investigate the holographic interpretation of composite defect CFTs. In \cite{Coccia:2021lpp}, the vacuum expectation value of a supersymmetric Wilson loop has been discussed in the context of AdS/BCFT correspondence (see also \cite{Gutperle:2020gez}). It is intriguing to investigate the holographic interpretation of various correlators in composite defect CFTs developed in this paper. In conventional defect CFTs, we can construct a scalar field which is located in an AdS spacetime from the defect OPE blocks \cite{Fukuda:2017cup}. It may be interesting to construct the dynamical degree of freedom (dof) in an AdS spacetime from the sub-defect OPE blocks. 
    \item \textbf{Monotonicity theorem for sub-defect entropy: }it is quite natural to speculate that the dof of a sub-defect should monotonically decrease along with the RG flow from a UV theory to an IR one. Therefore, we can expect the existence of the sub-defect entropy which captures the dof of a sub-defect. It is interesting to define this sub-defect entropy and prove the monotonicity theorem on it.

\end{itemize}

\acknowledgments
The author is particularly grateful to Tatsuma Nishioka for stimulating discussions on various points in this paper and encouraging the author to write this paper. The author is also grateful to Dongsheng Ge for discussions relating this project. The work of S.\,S. was supported by Grant-in-Aid for JSPS Fellows No.\,23KJ1533.

\appendix

\section{Method of images in composite defect CFTs}\label{sec: method of images scalar}
In the conventional defect CFTs, the method of images makes it quite helpful to construct defect conformal correlators \cite{Nishioka:2022ook}. From the symmetry-breaking structure \eqref{eq: residual symmetry}, we can convince ourselves that the method of images is also valid for fixing the form of correlators even in composite defect CFTs. The dictionary for scalar primaries is given as follows:
\begin{align}
\begin{aligned}
    &\left\langle\, \prod_{k=1}^{n}\CO_{\Delta_{k}}(X_{k})\, \prod_{k=1}^{m}\widehat{\CO}_{\widehat{\Delta}_{k}}(\widehat{Y}_{k})\, \prod_{k=1}^{s}\widecheck{\CO}_{\widecheck{\Delta}_{k}}(\widecheck{Y}_{k})\, \right\rangle \\
    &\qquad \approx 
    \left\langle\, \prod_{k=1}^{n}\CO_{\delta_{k}}(X_{k})\, \CO_{\delta_{k}}(X'_{k})\CO_{\delta_{k}}(\overline{X}_{k})\CO_{\delta_{k}}(\overline{X}'_{k})\prod_{k=1}^{m}\widehat{\CO}_{\widehat{\delta}_{k}}(\widehat{Y}_{k})\widehat{\CO}_{\widehat{\delta}_{k}}(\overline{\widehat{Y}}_{k})\, \prod_{k=1}^{s}\widecheck{\CO}_{\widecheck{\Delta}_{k}}(\widecheck{Y}_{k})\, \right\rangle_{\text{CFT}}\ , 
\end{aligned}
\end{align}
where the symbol $\approx$ means that the kinematical structures of both hand sides are equivalent to each other. Also, $\delta_{k}\equiv\Delta_{k}/4$ and $\widehat{\delta}_{k}\equiv\Delta_{k}/2$, and various coordinates are defined by
\begin{align}\label{eq: mirror config}
    \begin{aligned}
        X&\equiv (X^{A}\, ,\, X^{\alpha}\, ,\,X^{i})\qquad , \qquad X'\equiv (X^{A}\, ,\, X^{\alpha}\, ,\, -X^{i})\ , \\
        \overline{X}&\equiv (X^{A}\, ,\, -X^{\alpha}\, ,\,-X^{i})\qquad , \qquad
        \overline{X}'\equiv (X^{A}\, ,\, -X^{\alpha}\, ,\,X^{i})\ , \\
        \widehat{Y}&\equiv (Y^{A}\, ,\, Y^{\alpha}\, ,\,0)\qquad , \qquad
        \overline{\widehat{Y}}\equiv (Y^{A}\, ,\, -Y^{\alpha}\, ,\,0)\qquad , \qquad \widecheck{Y}\equiv (Y^{A}\, ,\, 0\, ,\,0)\ .
    \end{aligned}
\end{align}
As a sanity check, we consider the bulk one-point function $\langle\,\CO_{\Delta}(X) \, \rangle$. The above dictionary ensures that we can write this one-point function in terms of four-point one in a CFT:
\begin{align}
    \langle\, \CO_{\Delta}(X) \, \rangle \approx \langle\, \CO_{\delta}(X)\, \CO_{\delta}(X')\CO_{\delta}(\overline{X})\CO_{\delta}(\overline{X}')\,\rangle_{\text{CFT}}\ . 
\end{align}
We can then fix the bulk one-point function in the arm with the standard CFT knowledge:
\begin{align}
    \langle\, \CO_{\Delta}(X) \, \rangle \approx \frac{f(\xi)}{(X\circ X)^{\Delta/2}}\ , 
\end{align}
where $\xi$ is the cross-ratio which is defined by \eqref{eq: cross-ratio 1}. Remarkably, the two independent cross-ratios in CFT side become reduced to the single one $\xi$ after the restriction to the mirror configurations. We can similarly confirm that all examples treated in section \ref{subsubsec: Scalar correlators in composite defect CFT} are consistent with the method of images. 

\section{Proof for the selected operator expansion for free scalar fields.}\label{appendix: proof of OEs}
In section \ref{subsec: O(N) free scalar model}, we discussed various operator expansions of the fundamental O$(N)$ scalar field $\phi^{I}$. In this appendix, we present the proof for operator expansions by paying attention to \eqref{eq: OE for Neumann wrt defect}. To prove that this DOE is correct, it is enough to evaluate the bulk-defect two-point function $ \langle\, \phi^{I}(x)\, \widehat{\phi}^{\,J}(\hat{\bm{y}})\, \rangle$ in two ways. The first one has already been given in \eqref{eq: bulk_defect func Neumann defect}, and the second one is to directly substitute the DOE \eqref{eq: OE for Neumann wrt defect} into this two-point function. As a result of this plugging, the two-point function can be evaluated as follows:
\begin{align}
    \begin{aligned}
        \langle\, \phi^{I}(x)\, \widehat{\phi}^{\,J}(\hat{\bm{y}})\, \rangle&=
        \sum_{n=0}^{\infty}\frac{(-4)^{-n}}{n!\, \left(\frac{d-p}{2}\right)_{n}}\, \ |x_{\perp}|^{2n}\, \widehat{\bm{\Box}}_{p}^{n}\, \langle\, \widehat{\phi}^{\,I}(\hat{\bm{x}})\, \widehat{\phi}^{\,J}(\hat{\bm{y}})\, \rangle \\
        &=\delta^{IJ}\,\sum_{n=0}^{\infty}\frac{(-4)^{-n}}{n!\, \left(\frac{d-p}{2}\right)_{n}}\, \ |x_{\perp}|^{2n}\,  \widehat{\bm{\Box}}_{p}^{n}\,\left\{
        \frac{1}{|\hat{\bm{x}}-\hat{\bm{y}}|^{d-2}}\left[A+1+(B+C)\left(\hat{\eta}+1\right)^{-\frac{d-2}{2}}\right]\right\}\ ,\\
        &=\delta^{IJ}\,\sum_{n=0}^{\infty}\frac{(-4)^{-n}}{n!\, \left(\frac{d-p}{2}\right)_{n}}\, \ |x_{\perp}|^{2n}\,  \widehat{\bm{\Box}}_{p}^{n}\,\left\{
        \frac{A+1}{|\hat{\bm{x}}-\hat{\bm{y}}|^{d-2}}+\frac{B+C}{\left(|\hat{\bm{x}}-\hat{\bm{y}}|^2 +4\hat{x}\cdot \hat{y}\right)^{\frac{d-2}{2}}}\right\}\ ,
    \end{aligned}
\end{align}    
where in the second line, we used the explicit form of defect two-point function \eqref{eq: defect_defect func Neumann defect}. By using the following identities:
\begin{align}
    \begin{aligned}
        \widehat{\bm{\Box}}_{p}^{n}\, \frac{1}{|\hat{\bm{x}}-\hat{\bm{y}}|^{d-2}}&=4^{n}\left(\frac{d-2}{2}\right)_{n}\, \left(\frac{d-p}{2}\right)_{n}\frac{1}{|\hat{\bm{x}}-\hat{\bm{y}}|^{d-2+2n}}\ , \\
        \widehat{\bm{\Box}}_{p}^{n}\, \frac{1}{\left(|\hat{\bm{x}}-\hat{\bm{y}}|^2 +4\hat{x}\cdot \hat{y}\right)^{\frac{d-2}{2}}}&= 4^{n}\left(\frac{d-2}{2}\right)_{n}\, \left(\frac{d-p}{2}\right)_{n}\frac{1}{\left(|\hat{\bm{x}}-\hat{\bm{y}}|^2 +4\hat{x}\cdot \hat{y}\right)^{\frac{d-2}{2}+n}}\ , 
    \end{aligned}
\end{align}
the above summation can be reduced as follows:
\begin{align}
    \begin{aligned}
        \langle\, \phi^{I}(x)\, \widehat{\phi}^{\,J}(\hat{\bm{y}})\, \rangle&=\delta^{IJ}\,\frac{A+1}{|\hat{\bm{x}}-\hat{\bm{y}}|^{d-2}}\, \sum_{n=0}^{\infty}\frac{(-1)^{n}}{n!} \,\left(\frac{d-2}{2}\right)_{n}
            \frac{|x_{\perp}|^{2n}}{|\hat{\bm{x}}-\hat{\bm{y}}|^{2n}}\\
            &\qquad\qquad+\delta^{IJ}\,\frac{B+C}{\left(|\hat{\bm{x}}-\hat{\bm{y}}|^2 +4\hat{x}\cdot \hat{y}\right)^{\frac{d-2}{2}}}\, \sum_{n=0}^{\infty}\frac{(-1)^{n}}{n!} \,\left(\frac{d-2}{2}\right)_{n}
            \frac{|x_{\perp}|^{2n}}{\left(|\hat{\bm{x}}-\hat{\bm{y}}|^2 +4\hat{x}\cdot \hat{y}\right)^{n}}\ .
    \end{aligned}
\end{align}
We moreover employ the following identity on the summation:
\begin{align}
    \sum_{n=0}^{\infty}\frac{1}{n!} \,\left(\frac{d-2}{2}\right)_{n}\, (-\chi)^{n}=(1+\chi)^{-\frac{d-2}{2}}\ , 
\end{align}
and we immediately obtain the bulk-defect two-point function \eqref{eq: bulk_defect func Neumann defect}.
\bibliographystyle{JHEP}
\bibliography{DCFT}

\end{document}